\documentclass[pra,aps,superscriptaddress,twocolumn,letter,nopacs,nofootinbib,longbibliography]{revtex4-2}
\usepackage[english]{babel}
\usepackage[utf8]{inputenc}
\usepackage[T1]{fontenc}

\usepackage{amsthm}
\usepackage{amsmath}
\theoremstyle{plain} 
\usepackage{graphicx}
\usepackage[caption=false]{subfig} 
\usepackage{float} 
\usepackage[colorlinks=true, allcolors=blue]{hyperref}
\usepackage{braket}
\usepackage{bbm,amsfonts}
\usepackage[shortlabels]{enumitem}

\usepackage{mathtools}
\usepackage[title]{appendix}
\usepackage{dsfont} 
\usepackage{tikz}
\usetikzlibrary{quantikz2}
\usepackage{rotating} 
\usepackage{mathtools} 

\hypersetup{
	colorlinks=true,  
	linkcolor=blue,   
	citecolor=blue,   
	urlcolor=blue     
}

\newtheorem{theorem}{Theorem}

\newtheorem*{example*}{Example}

\newtheorem{definition}[]{Definition}

\newtheorem*{remark*}{Remark}
\newtheorem{corollary}[]{Corollary}

\usetikzlibrary{quantikz2}
\usepackage{circuitikz}

\usepackage{comment}

\DeclarePairedDelimiter\floor{\lfloor}{\rfloor}

\usepackage{tikz}
\usepackage{quantikz}
\usepackage{thmtools}

\newcommand{\bsc}{Barcelona Supercomputing Center, Plaça Eusebi G\"uell, 1-3, 08034 Barcelona, Spain}
\newcommand{\ub}{Universitat de Barcelona, 08007 Barcelona, Spain}
\newcommand{\ICFO}{ICFO-Institut de Ciencies Fotoniques, The Barcelona Institute of
Science and Technology, 08860 Castelldefels (Barcelona), Spain}
\newcommand{\UJ}{Institute of Theoretical Physics,
Jagiellonian University, Krakow, Poland}
\newcommand{\CFT}{Center for Theoretical Physics,
Polish Academy of Sciences, Warsaw, Poland}

\newcommand{\Tr}{\ensuremath{\operatorname{Tr}}}
\newcommand{\Fmax}{\ensuremath{F_{\rm{max}}}}
\newcommand{\Fexp}{\ensuremath{F_{\rm{exp}}}}
\newcommand{\rhoexp}{\ensuremath{\rho_{\rm{exp}}}}

\begin{document}

\title{Quantum Circuits for High-Dimensional Absolutely Maximally Entangled States} 
\author{Berta Casas}
\email{berta.casas@bsc.es}
\affiliation{\bsc}
\affiliation{\ub}
\author{Grzegorz Rajchel-Mieldzio\'c}
\affiliation{BEIT, ul.\ Mogilska 43, 31-545 Krak{\'o}w, Poland}
\author{Suhail Ahmad Rather}
\email{suhail@pks.mpg.de}
\affiliation{Max Planck Institute for the Physics of Complex Systems, 01187 Dresden, Germany}
\author{Marcin Płodzień}\affiliation{\ICFO}
\author{Wojciech Bruzda}\affiliation{\CFT}
\author{Alba Cervera-Lierta} \affiliation{\bsc}
\author{Karol {\.Z}yczkowski}\affiliation{\CFT}\affiliation{\UJ}


\begin{abstract} 
Absolutely maximally entangled (AME) states of multipartite quantum systems
exhibit maximal entanglement across all possible bipartitions. These states lead to  teleportation protocols that surpass standard teleportation schemes,
determine quantum error correction codes
and can be used 
to test  performance of current term quantum processors.
Several AME states can be constructed from graph states using minimal quantum resources. However, there exist other constructions that depart from the stabilizer formalism. In this work, we present explicit quantum circuits to generate exemplary non-stabilizer AME states
of four subsystems with four, six, and eight
levels each
and analyze their capabilities to perform quantum information tasks.
\end{abstract}

\maketitle

\section{Introduction}
 
Quantum correlations, such as entanglement and Bell correlations, are fundamental features of many-body quantum systems~\cite{horodecki2009entanglement,brunner2014bell,horodecki2024multipartiteentanglement}. Beyond their foundational significance, these phenomena are also recognized as essential resources for quantum technologies~\cite{RevModPhys.91.025001,https://doi.org/10.48550/arxiv.2405.05785}, namely quantum computing, communication, simulation, and sensing, which rely heavily on them. 

The primary technological approach to harnessing quantumness for quantum-enhanced tasks has largely focused on many-body systems with local dimension $d = 2$, i.e., qubits, particularly those exhibiting bipartite entanglement of the GHZ type. 
However, an open question remains regarding the technological utility of states with higher local dimensions $d > 2$, which offer richer entanglement structures. For example, while it is impossible to construct a quantum state of four qubits that is maximally entangled with respect to any bipartition, this limitation does not apply to systems with $d > 2$. Specifically, a special class of quantum states known as Absolutely Maximally Entangled (AME) states can achieve this property. AME states are characterized by maximal entanglement across all possible bipartitions of the system, ensuring that every subset of particles is maximally entangled with its complementary subset. This uniform distribution of entanglement makes AME states highly symmetric and robust.

AME states are very convenient for quantum information manipulation protocols  requiring high amount of entanglement between parties \cite{Goyeneche2014,Goyeneche2015}, including quantum teleportation, secret sharing \cite{Helwig2012} and fault-tolerant quantum computation \cite{Pastawski2015,Raissi2018,Mazurek2020}.
Since high entanglement has to be present in quantum devices to achieve any type of quantum advantage and since AME states are arguably the most entangled states that can be constructed, it is natural to propose the generation of AME states with quantum circuits as a demanding benchmark for quantum computers \cite{CerveraLierta2019}. 
 
Formally, an AME$(n, d)$ state is defined for a system of $n$ qudits with local dimension $d$, such that for any partition of the system into two subsets of equal size (or as equal as possible), the reduced density matrix of each subset is maximally mixed. 
The existence of AME states for any number of parties $n$ and any local dimension $d$ is an outstanding open problem in the field. 
While there are several methods to find AME states, particular attention is given to combinations of $(n,d)$ that cannot be classified with conventional mathematical machinery. Notable examples include the proof of the non-existence of an AME$(7,2)$ state, open for many years and solved in Ref.~\cite{Huber_2017}, or more recently the existence of the AME$(4,6)$ state~\cite{Rather_2022,Zyczkowski_2023}.

The aim of this work is to present explicit circuits to generate high-dimensional multipartite AME states, 
and to foster their experimental realization in platforms capable of implementing many-qudit states.  
In contrast to previous contributions \cite{CerveraLierta2019, Cieslinski_2023}, we present qudit quantum circuits that generate the recently discovered non-stabilizer AME states, i.e., AME states that neither have a graph state representation nor are locally equivalent to such states. While stabilizer states are efficiently simulable due to the Gottesman-Knill theorem, the non-stabilizer AME states we study exhibit entanglement structures beyond this regime. Their maximal entanglement, combined with a certain degree of non-Cliffordness, makes them promising candidates for benchmarking applications. Although AME graph states can be straightforwardly generated with quantum circuits, the states presented in this work require more sophisticated implementations.

We provide a robustness analysis of these AME states against noise by providing experimental fidelity benchmarks, which certify their multipartite high-dimensional entanglement and assess their applicability in teleportation protocols.
Given the progress in the preparation of many-qudit systems using platforms such as trapped ions~\cite{ringbauer2022universal} or photonic circuits~\cite{chi2022programmable}, we believe that the circuits provided in this work can be experimentally tested on high-dimensional quantum computing hardware.

This paper is organized as follows. First, we review the basic concepts and the construction of graph-state AME states proposed in the literature. In Sec.~\ref{sec:results}, we present the main results, consisting of the explicit circuits that generate the non-stabilizer states: AME$(4,4)$, AME$(4,6)$, and AME$(4,8)$. We use qudits of the appropriate dimensions and qubit systems that represent a high-dimensional qudit. Next, in Sec.~\ref{sec:experimental} we analyze the experimental guidelines to implement and certify these AME states. Moreover,  we analyze AME states for quantum information protocols such as quantum teleportation. Finally, the conclusions of this work are presented in Sec.~\ref{sec:conclusions}.

\section{Background}\label{sec:preliminaries}

In this section, we define the basic concepts used in this work to construct high-dimensional AME states with quantum circuits.

\begin{definition}[Absolutely Maximally Entangled (AME) State]
  A pure quantum state $|\psi\rangle$ of $n$ parties, each with local dimension $d$, is an Absolutely Maximally Entangled (AME) state if, for all possible bipartitions of the system into subsystems $A$ and $B$, such that $|A| = \lfloor n/2 \rfloor $, the reduced density matrix  
  \begin{equation}
      \rho_A = \operatorname{Tr}_B (|\psi\rangle\langle\psi|)
  \end{equation}is  maximally mixed. In other words, the reduced density matrix is proportional to the identity 
  \begin{equation}
      \rho_A = \frac{\mathbb{I}}{d^{|A|}}.
  \end{equation}
  To specify the number of particles $n$ and their local dimension $d$ in an AME state, we use the notation AME$(n,d)$. 
\end{definition}

\begin{figure}
    \centering
    \includegraphics[width=1\linewidth]{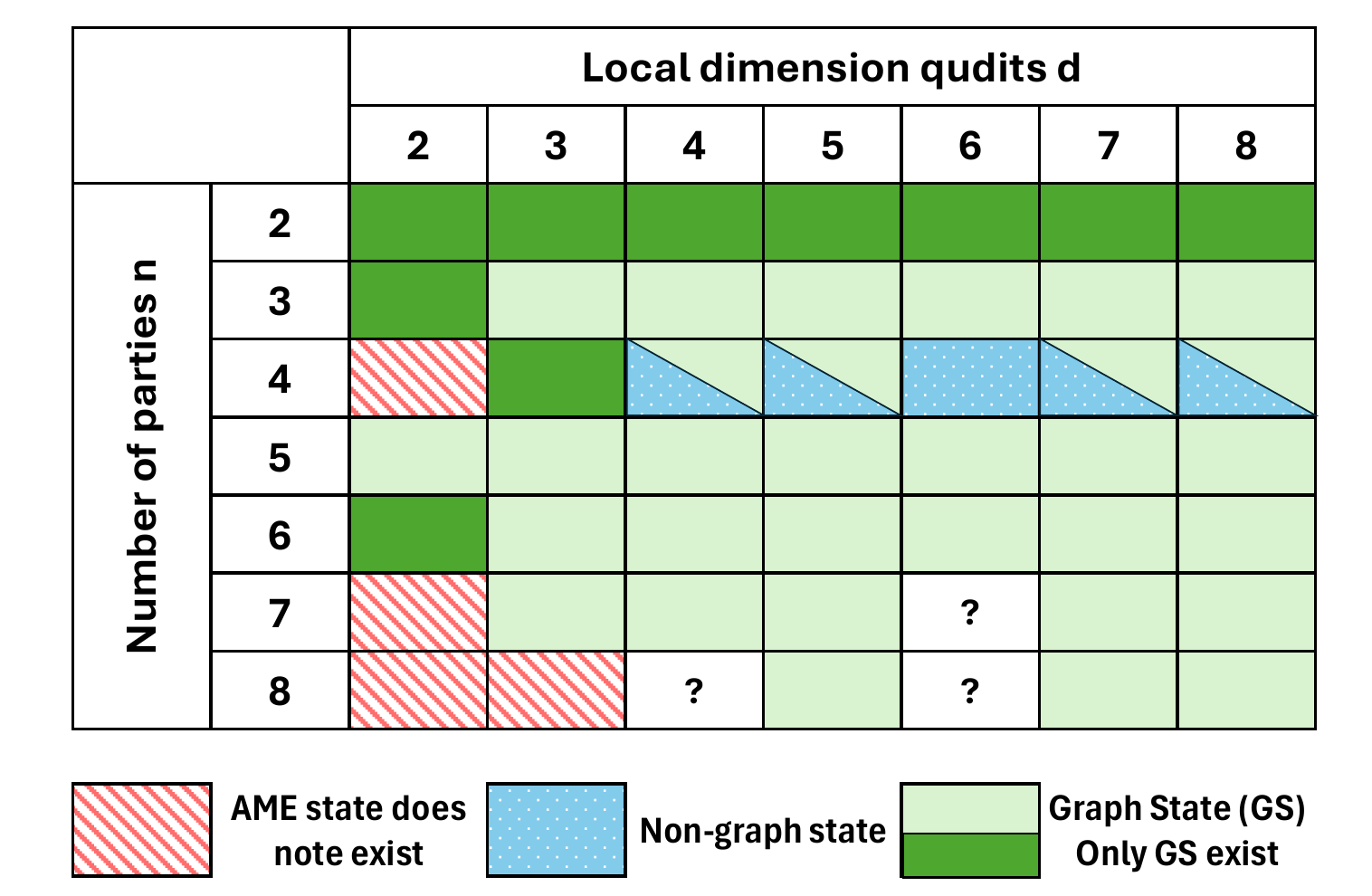}
    \caption{Table of known AME states up to local dimension $d=8$ and $n=8$ parties. Red entries indicate local dimensions and numbers of parties for which it has been proven that such an AME state does not exist. Blue entries represent cases where only non-graph AME states exist. Dark green denotes cases where only graph AME states exist, while light green indicates that a graph state construction is known, but other type of constructions are not excluded. For more details on additional dimensions, parties, specific construction, and references, see Ref.~\cite{TableAME}.} 
    \label{tab:AME}
\end{figure}

Defining multipartite entanglement classes becomes increasingly hard and ill-defined as the local dimension and number of parties grow. In that sense, several notions of entanglement exist. For the purpose of this work, we are interested in those states that remain maximally mixed regardless of which parties are traced out, i.e., AME states. An analytical way to test whether a state is AME is by computing the entropy of all bipartitions:

\begin{corollary}
    All bipartitions of an AME$(n,d)$ state have maximal Von Neumann entropy $S(\rho_{A})=-\rho_{A}\Tr(\rho_{A}) = \floor{n/2}$.
\end{corollary}

Most AME states can be represented as graph states. Table \ref{tab:AME} shows in green squares those AME states that admit this type of representation.
\begin{definition}[Qudit graph state ]
A graph $G$ is a pair $G = (V,E)$, where $V$ is set of $|V|$ nodes, while $E$ is a set of $|E|$ edges connecting any two nodes  $(j,k) \in E$, $j,k\in V$. 
A graph state $\ket{G}$ is defined on a graph $G$ as following: each node $j\in V$ corresponds to a qudit in the state $|+\rangle = \frac{1}{\sqrt{d}}\sum_{l=0}^{d-1}|l\rangle$ , while edge $(j,k)$ corresponds to the action of the $d$-dimensional controlled gate $CZ_d^{(jk)}$ between nodes $j$ and $k$,
\begin{equation}
    CZ_d^{(jk)}=\sum_{l,m=0}^{d-1}e^{\frac{2\pi i}{d}kl}|l\rangle_{j}\langle l|\otimes |m\rangle_{k}\langle m|.
    \label{eq:CZ_gate}
\end{equation}
Then, the qudit graph state takes the form
\begin{equation}
    |G\rangle = \left(\prod_{(j,k)\in E}CZ_d^{(jk)}\right)|+\rangle^{\otimes |V|}.
\end{equation}
\end{definition}
It is worth mentioning that, by construction, graph states always lead to stabilizer states.

From now on, we use $CZ_d$ to denote the controlled-$Z$ gate and drop the nodes $(jk)$ when understood from the context. From this definition, a graph state of $n$ qudits of dimension $d$ is constructed from an initial state $|+\rangle^{\otimes n}$, which can be constructed from the $|0\rangle^{\otimes n}$ state (the common initialization of any quantum computer) by applying the Fourier gate
\begin{equation}
    F_{d}=\frac{1}{\sqrt{d}}\sum_{k=0}^{d-1}\sum_{l=0}^{d-1}e^{\frac{2\pi i}{d}kl}|k\rangle\langle l|.
    \label{eq:Fourier_gate}
\end{equation}
For qubit graph states, this gate is the well-known Hadamard gate $H$, a standard quantum gate for current hardware. Therefore, the quantum circuit for preparing graph states admits a very simple decomposition.

\begin{figure}
\raggedright \hspace{0.36cm} $(a)$\\
    \centering 
    \includegraphics[width=0.8\columnwidth]{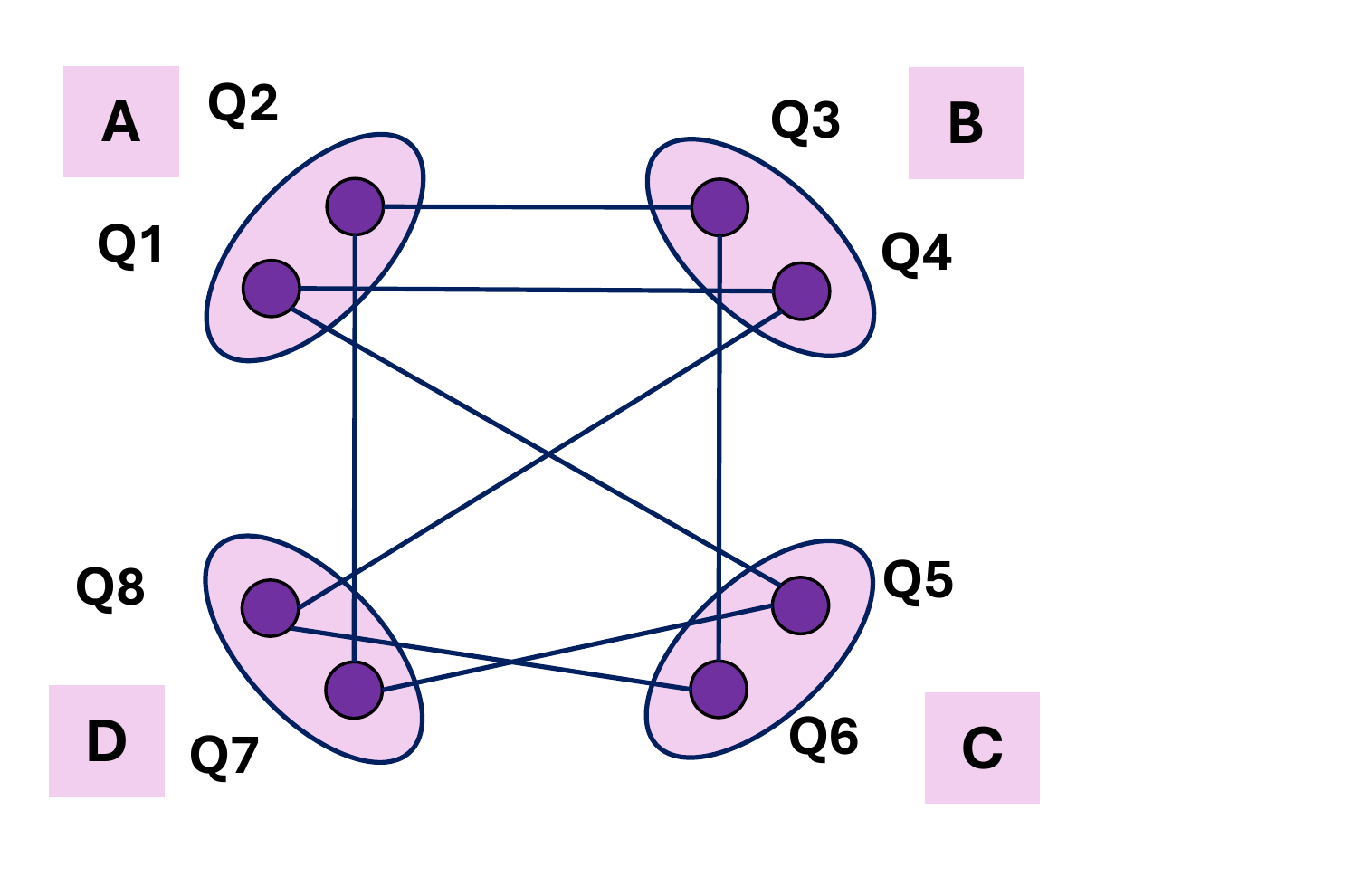}\\
    $(b)$
    \begin{quantikz}[row sep=0.1cm, column sep=0.5cm]
        \lstick[2]{A} Q1 \ \ket{0} & \gate{H} \slice{} & \qw & \ctrl{3} \slice{}& \ctrl{4} & \qw & \qw & \qw & \qw \\
         Q2 \ \ket{0} & \gate{H} & \ctrl{1} & \qw &  \qw &  \ctrl{5} & \qw & \qw & \qw\\
        \lstick[2]{B} Q3 \ \ket{0} & \gate{H} & \ctrl{-1} & \qw & \qw & \qw & \ctrl{3} & \qw & \qw\\
        Q4 \ \ket{0} & \gate{H} & \qw & \ctrl{-3} & \qw & \qw & \qw & \ctrl{4} & \qw\\
       \lstick[2]{C} Q5 \ \ket{0} & \gate{H} & \qw & \ctrl{2} & \ctrl{-4} & \qw & \qw & \qw & \qw\\
        Q6 \ \ket{0} & \gate{H} & \ctrl{2} & \qw & \qw & \qw & \ctrl{-3} & \qw & \qw\\
        \lstick[2]{D} Q7 \ \ket{0} & \gate{H} & \qw & \ctrl{-2}  & \qw & \ctrl{-5} & \qw & \qw & \qw\\
        Q8 \ \ket{0} & \gate{H} & \ctrl{-2} & \qw & \qw & \qw & \qw & \ctrl{-4} & \qw
    \end{quantikz}
    \caption{The state AME$(4,4)$ involves four parties $A,B,C$, and $D$, in which a ququart
    is represented by two qubits: $(a)$ the corresponding 8-qubit graph state borrowed from Ref.~\cite{helwig2013absolutely}; $(b)$ the circuit starting with Hadamard gates preparing the state $|+\rangle^{\otimes 8}$, followed by controlled-Z gates between qubits connected by edges in the graph. The circuit can be implemented in three steps (delineated by red dashed lines), with up to four gates applied in parallel.}
    \label{fig:AME44}
\end{figure}

Several AME states can be represented by graph states~\cite{helwig2013absolutely}.
As an example, Fig.~\ref{fig:AME44}$(b)$ shows the quantum circuit generated from the corresponding graph in Fig.~\ref{fig:AME44}$(a)$, where each node corresponds to a qubit in the state
$|+\rangle=H|0\rangle$, and each edge corresponds to a $CZ_2 \equiv CZ$, the two-qubit controlled-Z gate. For well-known two-qubit gates such as the Hadamard, controlled-Z and CNOT gates, we use the standard notation and drop the subscripts for convenience. 

Some data concerning the
graph representation of AME$(n,d)$ states
is provided in Tab.~\ref{tab:AME}. 
For further information on the existence of AME states for different number of parties
and local dimensions,
consult Ref.\cite{TableAME}.

In some cases it is convenient
to consider a qudit as a collection
of smaller subsystems.
For instance, the four-ququart state 
 AME$(4,4)$ can be interpreted as a 
 state of eight qubits system 
entangled in such a way that four
 parties, each consisting of 
two qubits, form an AME state -- see 
Fig.~\ref{fig:AME44}$(a)$.
 
 This example illustrates that AME states of certain dimension $d$ can be generated using lower-dimensional qudits of dimension $d'$ to represent each $d$-dimensional party. To achieve this, one needs to $(i)$ define a representation for the qudits of dimension $d$ and $(ii)$ determine the quantum gate decomposition of each $d-$dimensional operation using the $d'$-dimensional gates. As another example, in Ref.~\cite{CerveraLierta2019} the authors suggest using three computational basis states of two qubits to represent a qutrit ($|00\rangle\rightarrow |0\rangle$, $|01\rangle\rightarrow |1\rangle$, $|10\rangle\rightarrow |2\rangle$) and found the qubit gate representation of the AME$(4,3)$ qutrit gates. In other words, they simulate an AME$(4,3)$ state using $4\times2$ qubits.

\section{Results}\label{sec:results}

Under the operator-state mapping, pure quantum states of four-qudits each of local dimension $d$ are isomorphic to bipartite operators states on $\mathbb{C}^d \otimes  \mathbb{C}^d$. A pure state  $\ket{\mathcal{A}}$ in $\mathbb{C}^d \otimes \mathbb{C}^d  \otimes \mathbb{C}^d \otimes \mathbb{C}^d$ can be written in terms of a bipartite operator $\mathcal{A}$ as follows:
\begin{equation}
    \ket{\mathcal{A}}_{1234}:=(\mathcal{A}_{12} \otimes \mathbb{I}_{34}) \ket{\Phi}_{13} \otimes  \ket{\Phi}_{24},
    \label{eq:ket_4_party}
\end{equation}
where the bipartite state $\ket{\Phi}$ given by
\begin{equation}
    \ket{\Phi}:= \frac{1}{\sqrt{d}}\sum_{j=0}^{d-1} \ket{j}\ket{j},
\end{equation} 
is the generalized Bell state. The subscripts label the qudits, and note that the bipartite operator $\mathcal{A}$ acts on qudits $1$ and $2$, which share a Bell state with $3$ and $4$, respectively. 

The four-qudit state given in Eq.~(\ref{eq:ket_4_party}) is an AME state if all two-qudit reduced density matrices are maximally mixed. Considering the bipartitions: 12|34,13|24, 
and 14|23, it can be shown that the two-qudit reduced density matrices are given by
\begin{align}
\label{eq:A_R_Gam}
    \rho_{12} & =\Tr_{34}\left(\ket{\mathcal{A}}_{1234} \bra{\mathcal{A}}\right)=\frac{1}{d^2} \mathcal{A} \mathcal{A}^{\dagger},\\  \nonumber 
    \rho_{13} & =\Tr_{24}\left(\ket{\mathcal{A}}_{1234} \bra{\mathcal{A}}\right)=\frac{1}{d^2}\mathcal{A}^R \mathcal{A}^{R\,\dagger},\\
    \nonumber
    \rho_{14} & =\Tr_{23}\left(\ket{\mathcal{A}}_{1234} \bra{\mathcal{A}}\right)=\frac{1}{d^2}\mathcal{A}^{\Gamma} \mathcal{A}^{\Gamma \, \dagger},
\end{align}
where $\Tr_{ij}$ denotes the partial trace operation with respect to qudits $i$ and $j$. For $\mathcal{A}$ acting on $\mathbb{C}^d \otimes \mathbb{C}^d$, the  matrix rearrangements
$^R$ and $^\Gamma$, 
denoting reshuffling and partial
transpose, respectively,
are defined in a product basis,
\begin{align}
 \label{eq:U_R_Gam}
    \bra{k}\bra{l} \mathcal{A}^R \ket{i}\ket{j} & :=\bra{k}\bra{i} \mathcal{A} \ket{l}\ket{j},\\ \nonumber
    \bra{k}\bra{l} \mathcal{A}^{\Gamma} \ket{i}\ket{j} & :=\bra{k}\bra{j} \mathcal{A} \ket{i}\ket{l}. 
\end{align}
From the set of expressions in Eq.~(\ref{eq:A_R_Gam}), it follows that all two-qudit reduced matrices are maximally mixed, i.e.,  $\rho_{12}=\rho_{13}=\rho_{14}=\mathbb{I}/d^2$, if $\mathcal{A}$, $\mathcal{A}^R$, and $\mathcal{A}^{\Gamma}$ are all unitary. 
A bipartite unitary operator $\mathcal{U}$ for which $\mathcal{U}^R$ and $\mathcal{U}^{\Gamma}$ are also unitary is called 2-unitary \cite{Goyeneche2015,Rather_2022_Dual_Eq}. In this work we call such gates multi-unitary, as this terminology originally introduced in Ref.~\cite{Goyeneche2015} is now  commonly used in quantum many-body physics \cite{Bertini_2019_Exact,Jonay_2021_Tri_uni}.

\begin{figure}
    \centering    \includegraphics[width=1.0\linewidth]{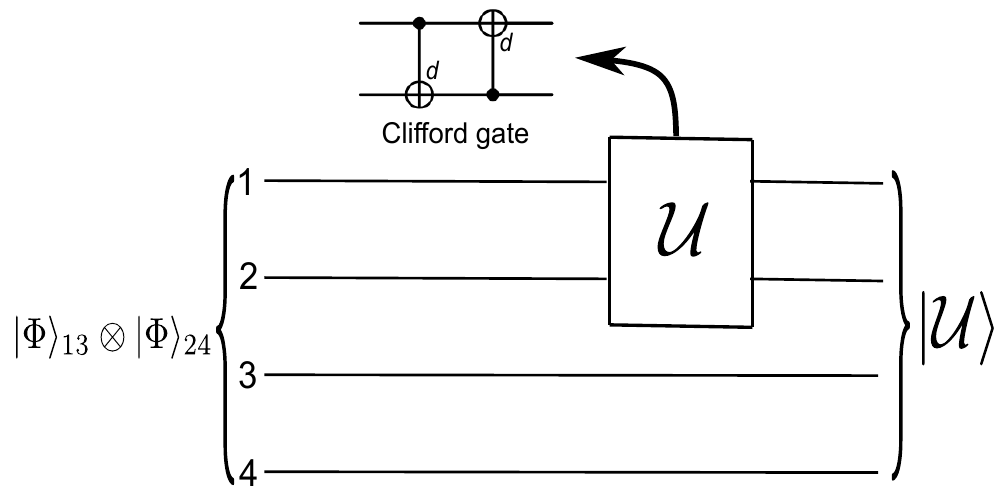}
    \caption{Schematic of operator-state mapping for an AME state of four qudits labeled from 1 to 4. (Left to right) A pair of Bell states is created: one Bell state $\ket{\Phi}_{13}$ (shared between qudits 1 and 3) and the other one $\ket{\Phi}_{24}$(shared between qudits 2 and 4). After the preparation of Bell states a ``multi-unitary'' gate $\mathcal{U}$ is applied between qudits 1 and 2. The output is an AME state of four qudits, denoted as $\ket{\mathcal{U}}$. Our main interest is in AME states for which the given multi-unitary gate can be written as a combination of simple entangling gates such as controlled gates as shown for stabilzer states in odd dimension $d$.}
    \label{fig:four_party}
\end{figure}

It is worth to mention that multi-unitary gates, when viewed as tensors, are called perfect tensors and have useful applications in holographic quantum error correction \cite{Pastawski2015}. In order to generate AME state experimentally, the corresponding multi-unitary should have a convenient quantum circuit representation as illustrated in Fig.~(\ref{fig:four_party}). For instance, the multi-unitary gates corresponding to graph states in odd dimensions are two-qudit Clifford gates and can be written as a simple concatenation of controlled-not gates \cite{Rather_2023}. A two-qudit Clifford gate $U$ maps a generalized Pauli group to itself under conjugation \cite{Hostens_20055_Stabilize}.

\subsection{Non-stabilizer four-qudit AME states}
Most of the known examples of AME states are stabilizer or graph states \cite{Helwig_2013_AME}. However, there exist cases in which constructions of stabilizer AME states either do not exist or are not known. A prominent example is the four-qudit system of six-level systems (quhexes). While no stabilizer AME states are known for four quhexes, explicit constructions of non-stabilizer AME states have been provided using various methods \cite{Rather_2022,Zyczkowski_2023,Rather_2023}.  In fact, it is strongly believed that no stabilizer AME$(4,6)$ state exists, though no formal proof has been established.

In order to distinguish AME states obtained from different constructions, it is useful to define the notion of local unitary (LU) equivalence. Two multipartite pure states $ \ket{\Psi}$ and  $\ket{\Psi'}$ of $n$ qudits are said to be LU equivalent if there exist single-qudit unitary gates $u_i \in \mathbb{U}(d)$ such that
\begin{equation}
    \ket{\Psi'}=\left(u_1\otimes u_2\otimes \cdots\otimes u_n\right) \ket{\Psi}.
    \label{eq:LUEq}
\end{equation}
Necessary and sufficient conditions for LU equivalence are known for bipartite quantum systems \cite{Nielsen_1999_Majorization}. The problem of LU equivalence among multipartite states is in general a difficult problem and many complementary methods have been developed to understand the structure and ordering of multipartite states \cite{Gour_2017_multipartite,Sauerwein_2018_LOCC}. 

The focus of our work is on AME states composed of four qudits for which the LU equivalence can be examined using the method of invariants of LU equivalence \cite{Rather_2023_AME_Eq}. Our main interest is in four-qudit AME states that are not LU equivalent to known AME states obtained from graph states. It is known that AME states of four qubits do not exist \cite{Higuchi_2000_AME42,Scott_2004_QECC,Huber_2018_AME_bounds}. Recently, it was shown in Ref.~\cite{Rather_2023_AME_Eq} that all AME states of four qutrits ($d=3$) can be transformed to a graph or stabilizer state using LU transformations. Therefore, $d=4$ is the smallest possible local Hilbert space dimension in which four-qudit AME states can exist that cannot be transformed into a graph state using LU transformations. We refer to such AME states as non-stabilizer AME states and provide explicit examples of such states. We emphasize that although the existence of such non-stabilizer AME states is already known in the literature \cite{Rather_2022,Rather_2022_Dual_Eq}, it is not known how to generate such states in practice using quantum circuits with qudit- and qubit-based encoding strategies. 

For non-stabilizer AME states, the corresponding multi-unitary gates $\mathcal{U}$ are not Clifford. There exist several numerical search algorithms to find such non-stabilizer AME states or, equivalently, non-Clifford multi-unitary gates \cite{Rather_2022_Dual_Eq}. Here we use the biunimodular construction discussed in Ref.~\cite{Rather_2023} to construct non-Clifford multi-unitary gates or perfect tensors. The main advantage of this method is that it provides a way to generate non-Clifford multi-unitary gates that are amenable to quantum circuit implementations as explained below.

Following Ref.~\cite{Rather_2023}, a class of AME$(4,d)$ states that is amenable to quantum circuit implementations can be obtained using special phase-valued (unimodular) $d\times d$ arrays or vectors called biunimodular arrays/vectors defined below.
\begin{definition}
    Biunimodular vector \cite{Fuhr_2015_Biuni}: A unimodular (phase-valued)  vector $\ket{\Lambda}$ of length $d$ is said to be biunimodular with respect to a unitary matrix $V$ in the unitary group $\mathbb{U}(d)$ if $V \ket{\Lambda}$ is also unimodular (phase-valued).
\end{definition}
Here we mainly work with biunimodular with respect to the Fourier matrices. Such vectors have a special property of vanishing periodic auto-correlations \cite{Fuhr_2015_Biuni}, which plays a crucial role in the construction of AME states \cite{Rather_2023}.

\begin{figure*}
    \centering
    \begin{quantikz}
        \ket{0} & \gate{F_d} \gategroup[4,steps=3,style={inner
sep=6pt, dashed}]{Bell state preparation} &  \ctrl[wire style={draw, "d"{pos=0.7, below, xshift=4.6pt, yshift=-5pt}}]{2} & &  &  & \ctrl[wire style={draw, "d"{pos=0.1, below, xshift=4.6pt, yshift=-5pt}}] {1}\gategroup[2,steps=5,style={inner
sep=6pt, dashed}]{Multi-unitary: $\mathcal U [\Lambda]$}  & \gate{F_d} & \gate[2]{D[\Lambda]} & \gate{F_d} &\ctrl[wire style={draw, "d"{pos=0.1, below, xshift=4.6pt, yshift=-5pt}}]{1} & \rstick[4]{$\ket{\mathcal{U}[\Lambda]}$}\\[-0.4cm]
        \ket{0} &\gate{F_d} & &\ctrl[wire style={draw, "d"{pos=0.7, below, xshift=4.6pt, yshift=-5pt}}]{2} &&   &\targ & & & & & \targ{} &\\[0.0cm]
        \ket{0} & & \targ{} & &  & & & & & & & \\[-0.2cm]
        \ket{0} & & & \targ{}  & &  & & & & & &
        \end{quantikz}
        \caption{General circuit for preparing an AME state of four particles with local dimension $d>3$. The Fourier gate of dimension $d$ is denoted as $F_d$, defined in Eq.~\eqref{eq:Fourier_gate}. The generalized control gate is $CX_d = \sum_{i=0}^{d-1} \ket{i}\bra{i} \otimes X^i$, where $X$ is the shift matrix of order $d$. The diagonal gate $D[\Lambda]$ of order $d^2$ depends on the specific AME state to prepare. Its diagonal entries form a biunimodular vector with respect to $F_d\otimes F_d$.}
\label{fig:AME(4,d)_circuit}
\end{figure*}
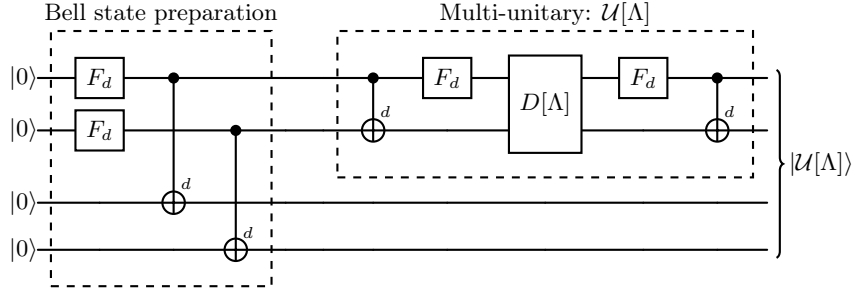   

For a given local dimension, let $\ket{\Lambda}$ be a biunimodular vector with respect to $F_d \otimes F_d$, where $F_d$ is the single-qudit Fourier gate defined in Eq.~(\ref{eq:Fourier_gate}).
Due to biunimodular property the unimodular entries of $\ket{\Lambda}$; $\lambda_{i,j}:=\bra{ij}\Lambda\rangle \in \mathbb{U}(1)$, have vanishing periodic auto-correlations:
\begin{equation}
    \sum_{k,l=0}^{d-1}\lambda_{i,j} \, \overline{\lambda}_{i\oplus_d k,j \oplus_d l}=0 ; \; (i,j)\neq(0,0),
    \label{eq:auto_corr}
\end{equation}
where $\overline{\lambda}$ denotes the complex conjugate of $\lambda$. A multi-unitary gate of size $d^2$ can be obtained from a biunimodular vector $\ket{\Lambda}$ if its unimodular entries satisfy the following additional properties, see Ref.~\cite{Rather_2023};
\begin{equation}
    \sum_{k,l=0}^{d-1} \omega_d^{(il-jk)} \,  \lambda_{i,j} \, \overline{\lambda}_{i\oplus_d k,j \oplus_d l}=0 ; \; (i,j)\neq(0,0),
    \label{eq:cross_auto_corr}
\end{equation}
where $\omega_d$ is $d$-th root of unity. Construction of unimodular vectors whose entries satisfy both Eq.~(\ref{eq:auto_corr}) and Eq.~(\ref{eq:cross_auto_corr}) is discussed in Ref.~\cite{Rather_2023} and such (bi) unimodular vectors exist for $d\geq 3$ including $d=6$. From such (bi) unimodular vectors a special class of multi-unitary gates can be constructed that are of the following form: 
\begin{equation}
    \mathcal{U}[\Lambda]=CX_d(F_d \otimes \mathbb{I})D[\Lambda](F_d^{\dagger} \otimes \mathbb{I}) CX_d^{T},
    \label{eq:U_biuni_gen}
\end{equation}
where $CX_d$ is the generalized controlled-NOT gate given by 
\begin{equation}
    CX_d = \sum_{j=0}^{d-1} |j\rangle\langle j| \otimes X_d^j.
    \label{eq:CNOT^k}
\end{equation}
The single-qudit gate $X_d$ is a generalization of the single-qubit Pauli-X matrix and is defined as $X_d |j\rangle = |j \oplus_d 1 \rangle$, where $\oplus_d$ denotes addition modulo $d$. The diagonal unitary $D[\Lambda]$ is obtained from the entries of $\ket{\Lambda}$,
\begin{equation}
D[\Lambda]=\sum_{i,j=0}^{d-1}\lambda_{i,j} |{ij}\rangle \langle{ij}|.
\end{equation}
The unitarity of the above multi-unitary gate with respect to $^R$ and $^\Gamma$ operations defined in Eq.~(\ref{eq:U_R_Gam}) results in Eq.~(\ref{eq:auto_corr}) and Eq.~(\ref{eq:cross_auto_corr}), respectively.

It is easy to verify that the controlled gates $CX_d$ and $CZ_d$ (see Eq.~\eqref{eq:CZ_gate} for the definition of $CZ_d$) are related as $CX_d=(\mathbb{I}\otimes F_d^{\dagger})CZ_d(\mathbb{I}\otimes F_d)$. We will use this relation to construct the circuits with a single Fourier gate.

\subsection{Circuits with qudits}
\label{sec:circuits with qudits}
Using the operator state mapping discussed above, a complete quantum circuit for generating an AME state using multi-unitary gates of the form given in Eq.~(\ref{eq:U_biuni_gen})
is shown in Fig.~\ref{fig:AME(4,d)_circuit}.
The circuit consists of two parts: 
\begin{itemize}
    \item Creation of a pair of (generalized) Bell states:
    \begin{equation}
    |\Phi\rangle_{13}\otimes |\Phi\rangle_{24} = \frac{1}{d}\sum_{i,j = 0}^{d-1} |ij\rangle_{12}\otimes|ij\rangle_{34}.
\end{equation}
    To create this state, all qudits are initialized in the $\ket{0}$ state, followed by the application of a Fourier gate on both the first and the second qudit. Next, two $CX_d$ gates are applied in parallel between qudits $(1,3)$ and $(2,4)$ resulting in a pair of generalized Bell states. 
    
\item Applying the multi-unitary gate: The remaining part of the circuit involves applying the multi-unitary gate $\mathcal U[\Lambda]$ given in Eq.~\eqref{eq:U_biuni_gen} between the first and second qudits.
\end{itemize}

For qudit-based quantum computing platforms, Fig.~\ref{fig:AME(4,d)_circuit} provides a way to prepare AME states for local dimensions $d\geq3$. Note that all the gates appearing in Fig.~\ref{fig:AME(4,d)_circuit} are well-known Clifford gates except the diagonal unitary $D[\Lambda]$. If the diagonal unitary is non-Clifford, the output of the quantum circuit is 
not a stabilizer AME state, and consequently, not a graph state either. One of the main challenges in experimentally preparing such non-stabilizer or non-graph states is decomposing the diagonal unitary in terms of gates from a universal qudit gate set.

Here, we propose a general method to construct arbitrary diagonal gates. In practice, the realization of such gates will depend on the native operations available in the chosen experimental platform, and these may need to be adapted accordingly. In what follows, we will discuss the experimental viability of these gates.

In general, a diagonal gate with arbitrary phases can be written as
\begin{equation}
D = \text{diag}(e^{ix_1}, e^{ix_2}, \dots, e^{ix_{d^2}}) := \text{diag}(e^{i\vec{x}}),
\end{equation}
where \(\vec{x} \in \mathbb{R}^{d^2}\) is an arbitrary real vector. In general, implementing such a gate is costly. We propose a decomposition strategy that partitions the $d^2$ diagonal elements into $d$ blocks: 
\begin{equation}
D = \sum_{j=0}^{d-1} |j\rangle \langle j| \otimes \text{diag} (e^{i\vec{x}_j}).
\end{equation}
In this decomposition, we divide \(\vec{x}\) into \(d\) sub-vectors,  
\begin{equation}
\vec{x}_j = (x_{jd}, x_{jd+1}, \dots, x_{jd+d-1}),
\end{equation}
where \(j \in \{0, \dots, d-1\}\), and each \(\vec{x}_j\) has dimension \(d\). This allows us to express the general diagonal gate as a sum of controlled operations.  

If the quantum hardware does not support simultaneous execution of all controlled operations, we can express the gate as a product of \(d\) controlled-phase operations:  
\begin{equation}
D = C_0(\vec{x}_0) C_1(\vec{x}_1) \dots C_{d-1}(\vec{x}_{d-1}),
\end{equation}
where each controlled operation is defined as  
\begin{equation}
C_l(\vec{x}_l) =  \sum_{\substack{j=0 \\ j\neq l}}^{d-1} |j\rangle \langle j| \otimes \mathbb{I} + |l\rangle \langle l| \otimes \text{diag} (e^{i\vec{x}_l}).
\end{equation}
The gate \(C_l(\vec{x}_l)\) acts as the identity on all elements except when the control qudit is in state \(|l\rangle\), in which case it applies a general diagonal phase gate on the target qudit.  

\subsection{Circuits with qubits}
Most of the currently available quantum computing platforms are based on qubits. 
Therefore, we provide a proper encoding of qudits into qubits and adapt the circuits discussed in the previous sections to use qubits gates. 
Taking into account the encoding of a qudit in terms of qubits, we modify Eq.~\eqref{eq:U_biuni_gen} into a form that is simpler for qubit-based quantum circuits, as explained below.

\subsubsection{AME state of four ququarts; AME$(4,4)$, using 8 qubits}

Assuming that each ququart ($d=4$) is encoded using two qubits, we need $4\times 2=8$ qubits in total to prepare a four-ququart AME state. One can construct such an AME state using a graph state, as shown in Fig.~\ref{fig:AME44}. Here, we present an alternative AME$(4,4)$ state that is not a graph state. Since the Fourier gate in a local dimension $d=4$; $F_4$, is not a tensor product of single-qubit Fourier gates, $F_4 \neq F_2 \otimes F_2$, it is important to modify Eq.~(\ref{eq:U_biuni_gen}) so that $F_4$ is replaced by the simpler gate $F_2 \otimes F_2$. Based on multi-unitary gates of the form given in Eq.~(\ref{eq:U_biuni_gen}), we consider an ansatz in which multi-unitary gates take the following form:
\begin{equation}
    \mathcal U[ \Lambda_{2,2}] = CZ_{2,2} (F_2 \otimes F_2)^{\otimes 2} D[\Lambda_{2,2}] (F_2 \otimes F_2)^{\otimes 2} CZ_{2,2},
    \label{eq:U_AME(4,4)_F2}
\end{equation}
where $F_2$ is the single-qubit Hadamard gate and $CZ_{2,2}$ is a controlled gate given by 
\begin{align*}
   & CZ_{2,2}  = 2 \sum_{k,l=0}^3 \left(\bra{l}F_2 \otimes F_2\ket{k}\right) \; |kl\rangle \langle kl|,\\
    & =\text{Diag}[1, 1, 1, 1, 1,-1, 1,-1, 1, 1,-1,-1, 1,-1,-1, 1].
\end{align*}
The subscript indicates that the diagonal entries (up to a scale factor) are elements of $F_2 \otimes F_2$.
The two-qudit unitary $D[\Lambda_{2,2}]$ is a diagonal unitary whose diagonal entries are obtained from a unimodular vector, $\ket{\Lambda_{2,2}}$, which is biunimodular with respect to $(F_2 \otimes F_2)^{\otimes 2}$ i.e., $\ket{\Lambda}$ and $(F_2 \otimes F_2)^{\otimes 2} \ket{\Lambda}$ are both unimodular. Such (bi) unimodular vectors also have vanishing periodic auto-correlations and satisfy equations similar to Eqs.~(\ref{eq:auto_corr}) and ~(\ref{eq:cross_auto_corr}). An example of such a biunimodular vector that leads to a multi-unitary gate is given by
\begin{equation}
   \ket{\Lambda_{2,2}}=[1,1,-i,i,i,i, 1, -1, i, -1, 1, i, i, -1, -1, -i ].
\end{equation}
This is obtained using a random search over unimodular vectors of length $d^2=16$ consisting of entries in a discrete group $\mathbb{G}_4=\left\lbrace\pm 1,\pm i\right\rbrace$. Substituting $D[\Lambda_{2,2}]$ obtained from the above biunimodular vector in Eq.~(\ref{eq:U_AME(4,4)_F2}), it can be verified by direct computation that  $\mathcal U[ \Lambda_{2,2}]$ is a multi-unitary gate i.e., it remains unitary under matrix arrangements $^R$ and $^{\Gamma}$ defined in Eq.~(\ref{eq:U_R_Gam}).  

Encoding each ququart into a pair of qubits, we decompose diagonal unitaries $CZ_{2,2}$ and  $D(\Lambda)$ appearing in Eq.~(\ref{eq:U_AME(4,4)_F2}) in terms of single- and two-qubit gates. The controlled gate $CZ_{2,2}$ viewed as a four-qubit gate has a simple transversal decomposition in terms of the two-qubit controlled-Z gate; $CZ_2 \equiv CZ$, given by
\begin{equation} 
\begin{quantikz} 
& \gate[4]{CZ_{2,2}} & \qw \\[-0.2cm]
 & & \qw \\ [0.2cm]
 & & \qw \\[-0.1cm]
 & & \qw \end{quantikz} = \begin{quantikz} & \qw & \qw & \ctrl{2} & \qw \\ & \qw & \ctrl{2} & \qw & \qw \\ & \qw & \qw & \control{} & \qw \\ & \qw & \control{} & \qw & \qw 
 \end{quantikz}.
 \label{eq:CZ_2_2_decom}
 \end{equation}
The other controlled gate $D[\Lambda_{2,2}]$ being non-Clifford does not have a simple decomposition in terms of two-qubit gates. Using the single-qubit phase gate; $S=\text{Diag}[1,i]$, two-qubit controlled-$S$ gate; $CS$, and the three-qubit double controlled-Z; $CCZ$, gate, we obtained the following decomposition for $D[\Lambda_{2,2}]$:
\begin{equation}
\begin{quantikz}
\label{eq:Dlam_2_2_decom}
&  \gate[4]{D[\Lambda_{2,2}]} &  \\[-0.1cm]
&  &   \\ [0.2cm]
&  &   \\ [-0.1cm]
&  &  
\end{quantikz} =
\begin{quantikz}
&\gate{S}&  \ctrl{1} & \ctrl{3} & \ctrl{2} & \ctrl{3} &\\[-0.3cm]
&\gate{S}& \gate{S^\dagger} &  & \control{}& &\\[-0.4cm]
&\gate{S}& \ctrl{1}&  & \control{}& \control{} &\\[-0.2cm]
&       & \control{} &  \gate{S}& & \control{} & \quad .\\[-0.3cm]
\end{quantikz}
\end{equation}

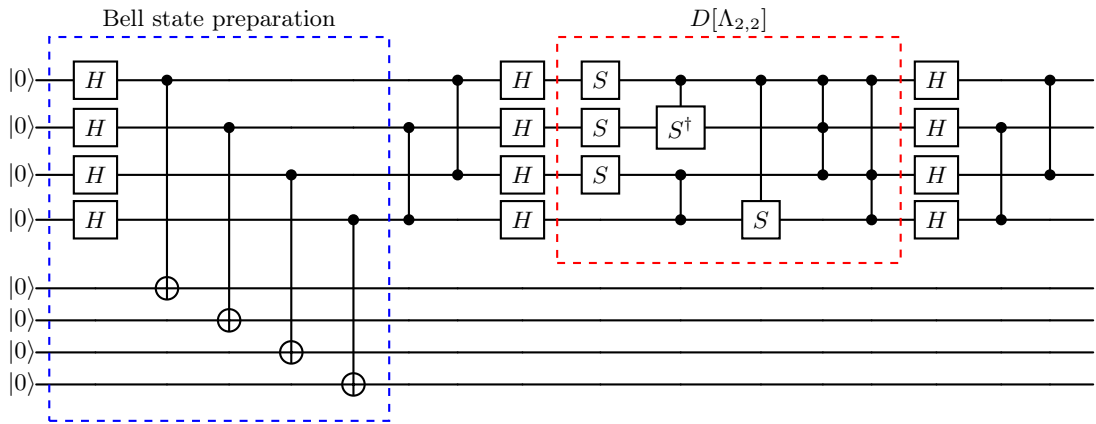
\begin{figure*}
    \centering
\begin{quantikz}
 \quad |0 \rangle & \gate{H} \gategroup[8,steps=5,style={inner
sep=6pt, dashed,color=blue}]{Bell state preparation} & \ctrl{4} &&&&&
\ctrl{2} & \gate{H} &\gate{S} \gategroup[4,steps=5,style={inner
sep=6pt, dashed,color=red}]{$D[\Lambda_{2,2}]$} &\ctrl{1}& \ctrl{3}& \ctrl{2}&\ctrl{3}& \gate{H} & & \ctrl{2}&   \quad  \\[-0.4cm]
 \quad  |0 \rangle &\gate{H}&&\ctrl{4}&&& 
\ctrl{2}&   & \gate{H}& \gate{S}& \gate{S^\dagger}&&\control{}& &\gate{H} & \ctrl{2} &  &  \quad  \\[-0.4cm]
 \quad  |0 \rangle &\gate{H}&&&\ctrl{4}&&&\control{}& \gate{H}& \gate{S}& \ctrl{1}&  & \control{}  & \control{}  & \gate{H}&& \control{} &  \quad  \\[-0.4cm]
 \quad  |0 \rangle &\gate{H}&&&&\ctrl{4}&\control{}&     & \gate{H} &  & \control{}&  \gate{S}  &  &\control{}  & \gate{H}& \control{}&  &  \quad  \\
 \quad  |0 \rangle & & \targ{}  & &  & & & & & & & &&&&&&  \quad  \\[-0.4cm]
 \quad  |0 \rangle & &  & \targ{} & & &  & & & & & &&&&&&  \quad \\[-0.4cm]
 \quad  |0 \rangle & & &  & \targ{} & & & & & & & &&&&&&  \quad \\[-0.4cm]
 \quad  |0 \rangle & & &  &  & \targ{} & & & & & & &&&&&&  \quad 
\end{quantikz} 
    \caption{Eight-qubit quantum circuit for an AME state of four ququarts; AME$(4,4)$. Each ququart ($d=4$) is encoded into two qubits and four Bell states are created using four transversal two-qubit CNOT gates along with four single-qubit Hadamard; $H \equiv F_2$, gates. Then a multi-unitary gate; $\mathcal{U}[\Lambda_{2,2}]$, given by Eq.~(\ref{eq:U_AME(4,4)_F2}) is applied on the first four-qubits. The multi-unitary gate; $\mathcal{U}[\Lambda_{2,2}]$, is decomposed using single and multi-qubit gates [Cf. Eqs.~(\ref{eq:CZ_2_2_decom}) and (\ref{eq:Dlam_2_2_decom})]. Note that apart from the familiar single- and two-qubit Clifford gates; Hadamard gate ($H$), phase-gate ($S$), CNOT gate ($CX_2$), and controlled-Z gate ($CZ_2$), the circuit also contains non-Clifford three-qubit double controlled-Z ($CCZ$) gate.
    }
    \label{fig:AME4_4_F_2_2}
\end{figure*}

Using Eqs.~(\ref{eq:CZ_2_2_decom}) and (\ref{eq:Dlam_2_2_decom}) in Eq.~(\ref{eq:U_AME(4,4)_F2}), a complete 8-qubit quantum circuit that results in a non-stabilizer AME$(4,4)$ state is shown in Fig.~\ref{fig:AME4_4_F_2_2}. Another representation of this state, in which gates  more evenly distributed across all qubits throughout the circuit, is discussed in Appendix \ref{app:AME_4_4_8_qubit_circ}.

\subsubsection{AME state of four quhexes; AME$(4,6)$, using $12$ qubits}
\label{sec:AME46_qubits}
We now discuss the construction of four-qudit AME states in local dimension $d=6$. Unlike other local dimensions, there is no known graph state in $d=6$ that is an AME state. In section~\ref{sec:circuits with qudits} we provided explicit quantum circuits of AME$(4,6)$ states with qudits of dimension six. Since $d=6=2\times 3$, we provide a circuit construction that can be implemented on a mixed quantum computing hardware architecture consisting of qubits and qutrits.

Similar to the ansatz in Eq.~(\ref{eq:U_AME(4,4)_F2}), for multi-unitary gates in local dimension $d=6$ we also assume the following form:
\begin{equation}
    \mathcal{U}[\Lambda_{2,3}]= CZ_{2,3} (F_2 \otimes F_3)^{\otimes 2} D[\Lambda_{2,3}] (F_2 \otimes F_3)^{\otimes 2} CZ_{2,3},\hspace{-.5em}
    \label{eq:U_AME_4_6_F_23}
\end{equation}
where $F_2\equiv H$ ($F_3$) is the single-qubit (qutrit) Hadamard gate and $CZ_{2,3}$ is a controlled gate given by 
\begin{align*}
    CZ_{2,3} & = \sqrt{6} \sum_{k,l=0}^5 (\bra{l}F_2 \otimes F_3\ket{k}) \; |kl\rangle \langle kl|.
\end{align*}
The unitary $D[\Lambda_{2,3}]$ is a diagonal unitary whose diagonal entries form a biunimodular vector of length $d^2=36$ with respect to $(F_2\otimes F_3)^{\otimes 2}$. Unlike in $d^2=16$ case, the random search algorithm is not efficient to find biunimodular vectors in $d^2=36$. Therefore, we use an iterative numerical algorithm discussed in Ref.~\cite{Rather_2023} for generating biunimodular vectors. An example of a simple biunimodular vector (with respect to $(F_2\otimes F_3)^{\otimes 2}$) obtained from the iterative numerical algorithm is given by
\begin{widetext}
\begin{equation}
\ket{\Lambda_{2,3}}=[ 1, \overline{\omega_3}, \overline{\omega_3}, \overline{\omega_3}, \overline{\omega_3}, 1, 1, 1, \omega_3, 1, \omega_3, 1, 1, \omega_3, 1, \omega_3
                , 1, 1, 1, \omega_3, 1, \overline{\omega_3}, \omega_3, \overline{\omega_3}, \overline{\omega_3}, \omega_3, \omega_3, \overline{\omega_3}, 1, 1, \omega_3, \omega_3, \overline{\omega_3}, 1, 1, \overline{\omega_3}],
\label{eq:diagonal_gate_qubits}
\end{equation}
\end{widetext}
where $\omega_3=\exp(2 \pi i/3)$ and $\overline{\omega_3}$ is the complex conjugate of $\omega_3$. The biunimodular vectors obtained numerically usually have a large number of distinct entries and it may be difficult to write an exact form. However, we focus only on those biunimodular vectors that have a small number of distinct entries, possibly, belonging to a (discrete) group. Note that the above biunimodular vector of length 36 has only 3 distinct entries belonging to the discrete group; $\mathbb{G}_3=\left\lbrace 1,\omega_3,\overline{\omega_3} \right\rbrace$. Substituting $D[\Lambda_{2,3}]$ obtained from the above (bi) unimodular vector in Eq.~(\ref{eq:U_AME_4_6_F_23}) leads to a multi-unitary gate.

This construction of the AME$(4,6)$ state is suited to be implemented with a mixed hardware architecture,  where each qudit of dimension six is represented by a qubit and a qutrit, which exactly matches the local dimension of each particle. We also provide a decomposition of the gate in Eq.~\eqref{eq:diagonal_gate_qubits} with qubits. For more details, we refer the reader to App.~\ref{app:qubit_circuit}.

It is worth mentioning that the above examples are not the only possible AME$(4,6)$ states. Recently, many other representatives have been proposed,
in particular the one with an unexpectedly rich internal structure, featuring additional degrees of freedom that potentially allow for greater flexibility~\cite{BZ24}. Nevertheless, our primary goal is to present the simplest possible scheme that will facilitate its experimental realization by avoiding the sequence of quantum gates required to reconstruct the large diagonal matrix. To the best of our knowledge, the constructions presented above appear to offer the most promising and straightforward candidates,
as it is currently unknown what the simplest form of such a state could be.

\subsubsection{AME state of four quoct; AME$(4,8)$, using $12$ qubit quantum circuit
}\label{sec:ame(4,8)}

\begin{figure*}
\begin{quantikz}
\ket{0} & \gate{H}  & \ctrl{6} & & & & & & \ctrl{3} & & & \gate{H} & \gate[6]{D[\Lambda_{2,2,2}]} & \gate{H} & \ctrl{3} & & & \\[-0.4cm]
\ket{0} & \gate{H} & & \ctrl{6} & & & & & & \ctrl{3} & & \gate{H} & &  \gate{H} & & \ctrl{3} & & \\[-0.4cm]
\ket{0} & \gate{H} & & & \ctrl{6} & & & & & & \ctrl{3} & \gate{H} & &  \gate{H} & & & \ctrl{3} &  \\[-0.3cm]
\ket{0} & \gate{H} & & & & \ctrl{6} & & & \control{} & & & \gate{H} & & \gate{H} & \control{} & & & \\[-0.4cm]
\ket{0} & \gate{H} & & & & & \ctrl{6} & & & \control{} & & \gate{H} & &  \gate{H} & & \control{} & & \\[-0.4cm]
\ket{0} & \gate{H} & & & & & & \ctrl{6} & & & \control{} & \gate{H} & &  \gate{H} & & & \control{} &  \\
\ket{0} & & \targ{} & & & & & & & & & & & &  & & & \\[-0.4cm]
\ket{0} & & & \targ{} & & & & & & & & & & &  & & & \\[-0.4cm]
\ket{0} & & & & \targ{} & & & & & & & & & &  & & &  \\[-0.4cm]
\ket{0} & & & & & \targ{} & & & & & & & & &  & & & \\[-0.4cm]
\ket{0} & & & & & & \targ{} & & & & & & & & & & &  \\[-0.4cm]
\ket{0} & & & & & & & \targ{}  & & & & & & &  & & & \\
\end{quantikz}    
\caption{Twelve-qubit quantum circuit for an AME state of four quocts; AME$(4,8)$. Each quoct ($d=8$) is encoded into three qubits and six Bell states are created using six transversal two-qubit CNOT gates along with six single-qubit Hadamard gates. Then a multi-unitary gate; $\mathcal{U}[\Lambda_{2,2.2}]$, given by Eq.~(\ref{eq:U_AME_4_8_F_2}) is applied on the first six-qubits. The main challenge is to decompose the six-qubit diagonal unitary $D[\Lambda_{2,2.2}]$ obtained from Eq.~(\ref{eq:D_lam_2_2_2}) in terms of single- and two-qubit gates.}
\label{fig:AME_4_8_circ}
\end{figure*}
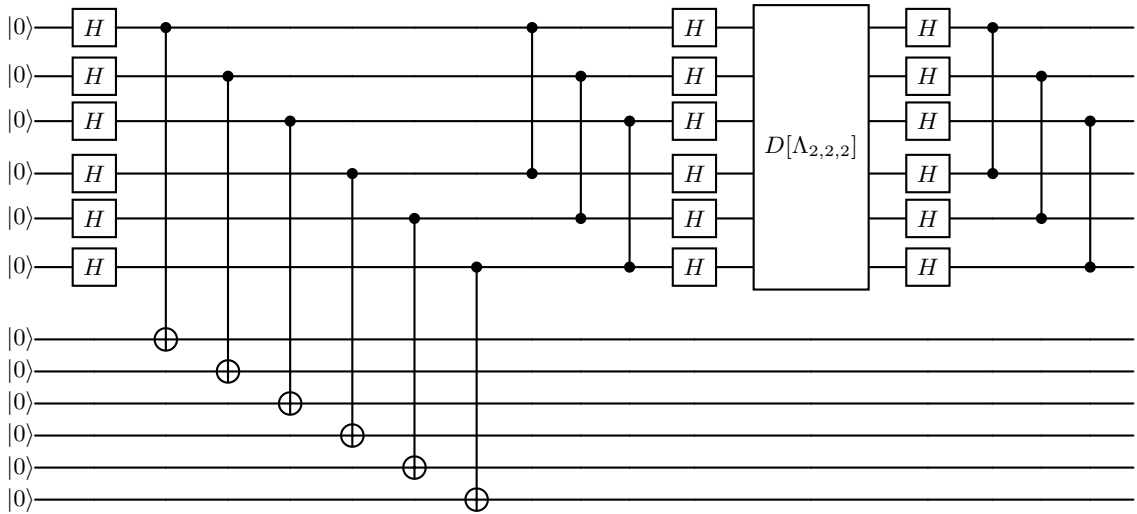

The local dimension $d=8=2^3$ is well-suited for qubit encoding as $\mathbb{C}^8 \cong \mathbb{C}^2 \otimes \mathbb{C}^2 \otimes \mathbb{C}^2$. Here we provide the construction of AME$(4,8)$ states together with their qubit-based quantum circuits.
Consider an ansatz for multi-unitary gates of the following form:
\begin{multline}
    \mathcal{U}[\Lambda_{2,2,2}] = CZ_{2,2,2} (F_2\otimes F_2\otimes F_2)^{\otimes 2}  \\
    D[\Lambda_{2,2,2}] (F_2\otimes F_2\otimes F_2)^{\otimes 2} CZ_{2,2,2},
    \label{eq:U_AME_4_8_F_2}
\end{multline}
where $F_2\equiv H$ is the single-qubit Hadamard gate and $CZ_{2,2,2}$ is a controlled gate given by 
\begin{align*}
    CZ_{2,2,2} & = \sqrt{8} \sum_{k,l=0}^7 (\bra{l}F_2 \otimes F_2 \otimes F_2\ket{k}) \; |kl\rangle \langle kl|.
\end{align*}
The unitary $D[\Lambda_{2,2,2}]$ is a diagonal unitary whose diagonal entries form a biunimodular vector of length $d^2=64$ with respect to $(F_2\otimes F_2\otimes F_2)^{\otimes 2}$. As in the previous case we use the numerical search algorithm to obtain biunimodular vectors with respect to $(F_2\otimes F_2\otimes F_2)^{\otimes 2}$. An example of such a biunimodular vector of length $d^2=64$ that leads to a multi-unitary gate is 
\begin{widetext}
\begin{align}
 \label{eq:D_lam_2_2_2}
\ket{\Lambda_{2,2,2}}= [& +,  +,  i,  +, -i,  -,   i,  +,  -i,  -,  -, -, i, -i,  +,   -, 
                             i,  i, -i, -i,  +, -i,  -i,  i,  -i,  +,  -, i, -,  +,  i,   -, \\ \nonumber
                           & -,  +, -,   i,  -,  +,   i, -i,   +,  +, -i, -, -,  +,  i,  -i, 
                             -,  -, -,   -,  -,  +,  -i,  +,   -,  -,  -, -,  i, i, -i,  -],
\end{align}
\end{widetext}
where $\mp$'s denote $\mp 1$'s to increase readability. Substituting $D[\Lambda_{2,2,2}]$ obtained from the above (bi) unimodular vector in Eq.~\eqref{eq:U_AME_4_8_F_2} leads to a multi-unitary gate $\mathcal U[\Lambda_{2,2,2}]$. A 12-qubit quantum circuit that results in an AME$(4,8)$ state using the multi-unitary gate given by Eq.~\eqref{eq:U_AME_4_8_F_2} is shown in Fig.~(\ref{fig:AME_4_8_circ}). The decomposition of the diagonal gate $D[\Lambda_{2,2,2}]$ in terms of single- and two-qubit gates is not simple but can be done using the Qiskit software \cite{qiskit2024} and is given in Appendix \ref{app:Diag_2_2_2_qubit_decom}.

The AME states obtained using the biunimodular construction in local dimensions $d=4,6,$ and $8$ are not local unitary (LU) equivalent to stabilizer or graph states. This is discussed in the Appendix \ref{app:LU_Invariants} using invariants of LU equivalence. 

\vfill

\section{Experimental guidelines}\label{sec:experimental}

\subsection{Qudit quantum gates}
To construct AME state circuits with qudit experimental platforms, a specific subset of quantum gates is required. One fundamental gate is the Fourier gate defined in Eq.~\eqref{eq:Fourier_gate}, a single-qudit operation that can be implemented in various quantum computing platforms. In photonic systems, it is naturally realized using linear optical components such as beam splitters and phase shifters~\cite{Chi_2022}. In superconducting circuits, is implemented using techniques that leverage the multi-level structure of superconducting qudits, while in trapped-ion systems, the Fourier gate can be engineered via Raman transitions or composite pulse sequences~\cite{ringbauer2022universal}.

AME states derived from graph states require the Fourier gate and the generalized Controlled-Z gate from Eq.~\eqref{eq:CZ_gate} ~\cite{ringbauer2022universal}. Other AME constructions additionally rely on the generalized controlled-not gate $CX_d$ and a diagonal gate $D(\Lambda)$, which depends on the specific AME state. Furthermore, depending on the biunimodular vector considered, the $CX_d$ gate may be replaced by another controlled operation (see, for example, Section~\ref{sec:AME46_qubits}).

The decomposition of diagonal gates $D(\Lambda)$ into controlled-phase operations between different levels is detailed in Section~\ref{sec:circuits with qudits}. If arbitrary controlled-level operations with variable phases are available, six controlled operations are sufficient to implement the required quantum gate. In practice, the decomposition of this gate must be adapted to the available entangling operation in the native hardware. For instance, a trapped-ion-based programmable qudit quantum computer~\cite{ringbauer2022universal} has demonstrated control over up to seven internal levels, benchmarking single-qudit and two-qudit gates, including the high-dimensional CNOT and M{\o}lmer-S{\o}rensen gates, which can be used to construct controlled-phase and Controlled-Z operations. Photonic~\cite{chi2022programmable} and superconducting circuit platforms~\cite{cervera2022experimental,goss2022high} also provide qudit control, though scalability beyond four levels and three-party interactions remains challenging. At present, trapped ions appear to be the most promising platform for implementing high-dimensional AME states.

\subsection{Multipartite high-dimensional entanglement certification}
In this section, we explore the minimal experimental fidelity required to ensure genuine multipartite entanglement in AME states, following the methods developed in Ref~.\cite{malik2016multi}. Specifically, we derive the fidelity threshold for an AME$(4,d)$ state that ensures the state being genuinely entangled in a $d^4$-dimensional space. This implies that the entanglement structure of the state cannot be understood as lower-dimensional entangled states. The results are summarized in the following theorem:
\begin{theorem}
\label{th:genuine_entanglement}
    Consider an AME state consisting of $4$ particles, each with dimension $d\in \mathds N$ and $d \geq 3$. The state is guaranteed to be genuinely multipartite entangled in a space of dimension $d^4$ if the fidelity of an experimental state with respect to the ideal state satisfies
    \begin{equation}
        \Fexp^{(4,d)} \geq \frac{d-1}{d}
    \end{equation}
\end{theorem}
We prove this result in Appendix~\ref{app:entanglement}. The requirement of having genuine multipartite entanglement across all dimensions becomes increasingly challenging when considering AMEs of higher local dimensions. For the specific case of an AME state with four particles, each of local dimension $d = 6$, the fidelity threshold is $\Fexp^{(4,6)}\geq 5/6 \approx 0.83$. Achieving this fidelity may be challenging for certain experimental setups, given the complexity of the circuit required to prepare such a state. However, in the following sections, we argue that achieving genuine multipartite entanglement in a $4^d$-dimensional space may exceed the requirements for some practical applications.

\subsection{Noise robustness high-dimensional AME states}
To implement the AME states on a physical architecture, it is necessary to understand the acceptable noise level on the total state. 
To this end, let us study the effect of the depolarizing noise on the entanglement properties across all bipartitions. Depolarizing noise is a suitable model for most architectures, as it captures the loss of coherence and mixing with the maximally mixed state, which is common in various quantum platforms.
The level of noise is quantified via a parameter $\gamma\in[0,1]$:
\begin{equation}
    \rho_\text{noisy} = (1 - \gamma) \ket{\psi}\!\!\bra{\psi} + \gamma \,\frac{\mathbb{I}}{6^4},
\end{equation}
where $\ket{\psi}$ is the pure state of our choice. 
Note that such a description of a state is equivalent to a channel of depolarizing noise. 

As a measure of entanglement that is straightforward to compute and suitable for both pure and mixed states, we choose negativity -- the sum of negative eigenvalues $\lambda_i$ of the partially transposed state $\rho^{\Gamma}$
\begin{equation}
    \mathcal{N}(\rho) = \sum_{\lambda_i < 0} |\lambda_i|.
\end{equation} 
There are three possible balanced bipartitions, so we take the sum of negativities over all of them as the final figure of merit. 
The noise level $\gamma$ that removes all entanglement from the state is also an upper bound on the state’s robustness~\cite{Vidal_1999}.

\begin{figure}[h]
\includegraphics[width=\linewidth]{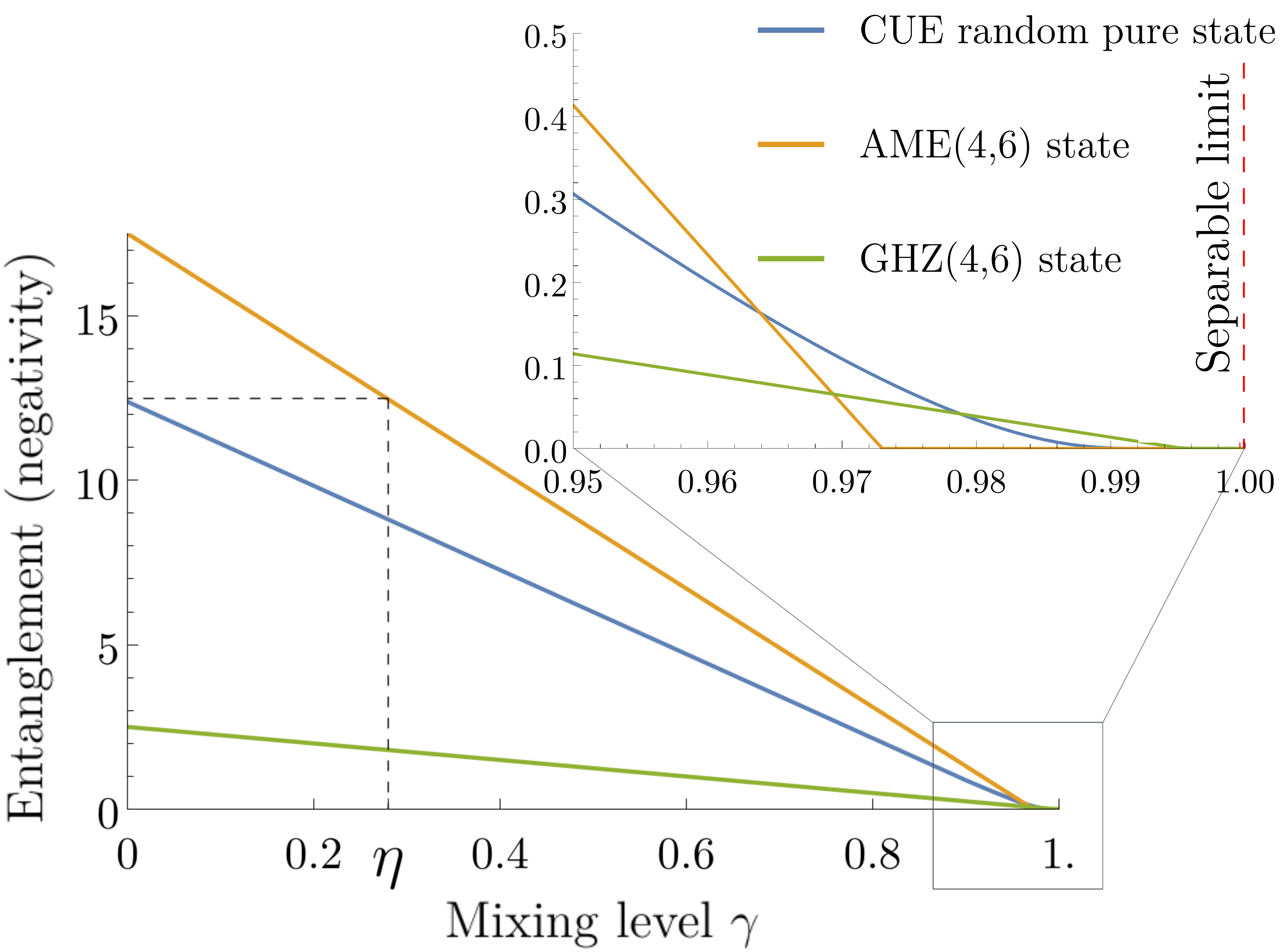}
	
    \caption{The decay of the negativity of entanglement as a function of the depolarizing noise parameter $\gamma\in[0,1]$ is analyzed for a generic state of four quhexes from the circular unitary ensemble (CUE), the golden AME$(4,6)$~\cite{Rather_2022} state, or the generalized GHZ state. 
    The value $\gamma = 0$ corresponds to pure states without noise, while for $\gamma = 1$ the resulting state is the maximally mixed state. 
    For noise level $\gamma$ smaller than $\eta = 28\%$, the AME$(4,6)$ state can be distinguished from a random state by its entanglement properties alone.
    A close-up of the region near the maximally mixed state shows that in this domain, CUE random pure states and GHZ states are more robust to mixing --there exists a set of noise parameters $\gamma$ for which the AME$(4,6)$ state is less entangled than its counterparts.
    The red dashed line delimits the region known to be separable for any starting pure state~\cite{Gurvits_2003}.
    We observed that all instances of the Haar-random pure state exhibited the same behavior, as the corresponding lines are indistinguishable.}
    \label{fig:robustness}
\end{figure}

In Fig.~\ref{fig:robustness}, we compare the loss of the entanglement quantified as the sum of negativities across balanced bipartitions with growing noise $\gamma$. 
We contrast AME$(4,6)$ state with two alternatives: generic state (Haar-random) and generalized GHZ state of four quhexes $|\text{GHZ}(4,6)\rangle$, where
\begin{equation}
    \ket{\text{GHZ}(n,d)} = \frac{1}{\sqrt{d}} \sum_{j = 0}^{d-1} \ket{j}^{\otimes n}.
\end{equation}
Up to a high level of noise, AME$(4,6)$ is superior to other states in terms of entanglement. 
Due to its relative robustness with respect to the noise, for $\gamma \leq 0.28$, AME$(4,6)$ can be differentiated from a noiseless Haar-random state.

AME states are understood as a different form of the generalized GHZ state.
In Fig.~\ref{fig:robustness}, it is possible to observe that their behaviour is quite different, as the GHZ state is not maximally entangled across all bipartitions. 
Indeed, the largest subsystems in the GHZ state that are maximally entangled are single particles -- the state is 1-uniform but not $2$-uniform. 
This means that in tasks that require higher-dimensional\footnote{Four-party GHZ state allows one
to teleport a state of a single subsystem of dimension $d$ to any other party, while any AME state
enables us to teleport any bipartite state of dimension $d^2$ to the remaining two parties.}
entanglement (two quhexes instead of one), it is necessary to use AME$(4,6)$ state. 

To conclude, for a level of noise $\gamma$ up to $\eta = 28\%$ AME$(4,6)$ state has higher entanglement than a generic pure state of the same dimensions. 
It is also much more entangled under this measure than the GHZ state. 
This can lead to a better estimation of an acceptable level of noise in the experimental setup ensuring that the state generated is truly better than a generic state. 
It is also of interest what ``generic'' means for a particular experiment, as a random state from the Hilbert space is highly entangled, even though this level of entanglement is hard to produce in practice. 
Taking into account the details of the setup might give an even better bound on the acceptable noise.

Let us conclude this subsection by discussing the requirements for single gates.
In order to properly evaluate acceptable noise, we add some realistic assumptions:
\begin{enumerate}
    \item $F_6$ and $P$ gates are much less noisy than $D(\Lambda)$.
    \item The preparation of the initial pair of Bell states $|\Phi\rangle_{13}\otimes |\Phi\rangle_{24}$ is also relatively noise-free.
    \item Noise is additive on the composition of gates.
\end{enumerate}
Using these assumptions as well as the decomposition of $D(\Lambda)$ into 6 controlled phase gates, we can estimate that the noise level of 4\% at the level of individual gates would reassure distinguishability from Haar-random gates.

\subsection{Teleportation protocol}
We showcase the capabilities of the noisy AME state on one of the most famous quantum information protocols.
Quantum teleportation is one of the most striking quantum techniques that do not have classical counterparts. 
The aim is to transfer the information in a quantum state from one place to another, without any physical translation of a particle. 
To achieve any quantum advantage, the use of entanglement is necessary. 
Therefore, AME states are perfect for teleportation, as they allow to transfer a state of $N/2$ particles from any bipartition to any other bipartition.

\begin{figure}[t!]
    \includegraphics[width=\linewidth]{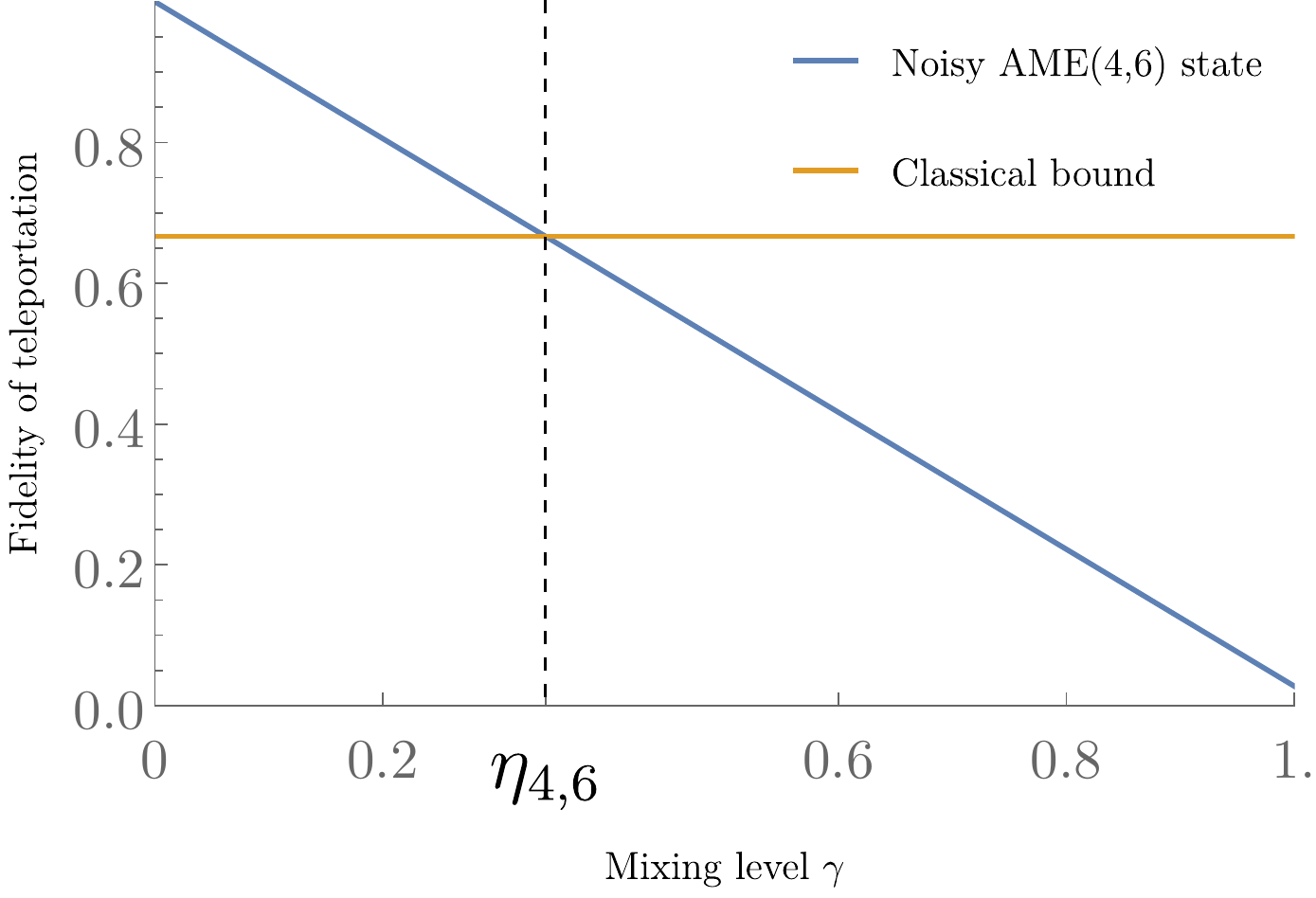}
    \caption{The fidelity of teleportation for 
    AME$(4,6)$ states under depolarizing noise with parameter $\gamma$.
    Up to noise level of $\eta_{4,6} = 12/35 \approx 34\%$, the protocol can achieve true quantum teleportation, beating classical fidelity of $2/3$.}
    \label{fig:teleportation_fidelity}
\end{figure}

As the noise inhibits all quantum protocols, similarly, there is a certain, acceptable level of noise for teleportation. 
For a maximally entangled state of dimension $d$, the fidelity between the input and the output state of the teleportation protocol that can be achieved reads~\cite{Fonseca_2019}
\begin{equation}
    \mathcal{F} = \frac{2}{d + 1}\bigg( 1 - \frac{d - 1}{2 d} \gamma\bigg) + \frac{d - 1}{d + 1} (1 - \gamma),
\end{equation}
for the depolarizing noise of parameter $\gamma$.
In particular, our findings concerning AME$(4,6)$ and AME$(4,4)$ states show that, up to noise levels of around $35\%$, both states even in the presence of noise are useful for teleportation, see Fig.~\ref{fig:teleportation_fidelity}. 
In those cases, $d = 16$ or $d=36$.

\section{Concluding remarks}\label{sec:conclusions}

After more than two decades of intensive analysis, the entanglement of multipartite quantum systems is understood relatively well. However, the experimental realization of several strongly entangled multipartite states remains a significant challenge. While GHZ states involving multiple qubits have been successfully constructed in the laboratory \cite{Hu11}, the {\sl Absolutely Maximally Entangled} (AME) states have yet to be achieved experimentally. These states,
related to quantum error correction codes,
exhibit quantum correlations among any selected subsystems
and are valuable for various quantum information processing protocols. As such, their construction in the laboratory 
offers applications for extended quantum
teleportation protocols and
serves as a practical benchmark for testing quantum hardware capabilities.

Reference \cite{CerveraLierta2019} proposed quantum circuits designed for experimental implementation on suitable quantum hardware, facilitating the construction of exemplary AME states. 
This raises an intriguing question: which  state is easier to construct in a lab, a six-qubit state AME$(6,2)$ belonging to a 
$64$-dimensional space
or a four-qutrit AME 
state from ${\cal H}_{81}$?

In this study, we extend the exploration beyond the commonly examined family of AME states associated with graph and stabilizer states. On one hand, we provide
an AME$(4,8)$ state, different from
the standard stabilizer states. 
The key result of this work - providing quantum circuits to generate exemplary AME states not belonging to this class,   
including the system composed of four subsystems with 
six levels each and related to the famous combinatorial problem of Euler \cite{Rather_2022,Zyczkowski_2023}, 
opens new avenues for the experimental fabrication of these highly non-classical states. Consequently, our work serves as a practical invitation to delve into {\sl quantum combinatorics}, a field initiated by the pioneering work of Zauner \cite{Zauner}: experimental construction of quantum analogues of classical combinatorial designs.

While AME states can be mathematically constructed and studied so far, there is no known physical process generating such states, or a known Hamiltonian whose ground state is an AME state.
This is in contrast to GHZ states, which can be realized as a ground state of the  Ising model. 
Therefore, it is currently not feasible to create AME states using analog quantum simulators, while our work pushes in the other direction, providing a construction of AME states on digital quantum simulators. 
From the perspective of experimentalists, AME states are harder to create than GHZ states.
However, AME states are \emph{the} most entangled states in the multipartite Hilbert space, with several times larger entanglement across all bipartitions than GHZ states even for four subsystems.

For these reasons, it would be a major technological breakthrough to realize AME states in the laboratory. 
While being hard to create, their entanglement is very robust to noise, as we have shown in this contribution. 
Furthermore, benchmarking these states is easily realizable in the laboratory, as one can measure R\'enyi entropy of order two, related to coincidence experiments, on a random bipartition~\cite{Liu_2024}.
First, the state is created, while later, the bipartition to check is chosen at random. 
We encourage future experimental efforts to construct the AME states researched in this paper, particularly in architectures that naturally allow for qudits, like trapped ions.

With the improvement of quantum computing hardware, high-dimensional quantum devices are emerging \cite{ringbauer2022universal, chi2022programmable,cervera2022experimental,goss2022high}, which makes it even more feasible to experimentally generate and use these AME states.
It will be fascinating to observe which existing experimental platforms prove to be most effective for this purpose.

\section*{Acknowledgements}
All the authors are thankful for the enlightening discussions with Dardo Goyeneche. B.C. thanks Jordi Tura and Eloïc Vallée for discussions on entanglement certification.
B.C., M.P. and A.C.-L. acknowledge funding from the Spanish Ministry for Digital Transformation and of Civil Service of the Spanish Government through the QUANTUM ENIA project call - Quantum Spain, EU through the Recovery, Transformation and Resilience Plan – NextGenerationEU within the framework of the Digital Spain 2026. A.C.-L. acknowledges funding from Grant RYC2022-037769-I funded by MICIU/AEI/10.13039/501100011033 and by “ESF+".
G.R.-M. acknowledges funding from the European Innovation Council accelerator grant COMFTQUA, no. 190183782.
M.P. acknowledges support from European Research Council AdG NOQIA;
MCIN/AEI (PGC2018-0910.13039/501100011033, CEX2019-000910-S/10.13039/501100011033, Plan National FIDEUA PID2019-106901GB-I00, Plan National STAMEENA PID2022-139099NB, I00, project funded by MCIN/AEI/10.13039/501100011033 and by the “European Union NextGenerationEU/PRTR" (PRTR-C17.I1), FPI); QUANTERA DYNAMITE PCI2022-132919, QuantERA II Programme co-funded by European Union’s Horizon 2020 program under Grant Agreement No 101017733. Fundació Cellex; Fundació Mir-Puig; Generalitat de Catalunya (European Social Fund FEDER and CERCA program. 
W.B. is supported by NCN SONATA BIS grant no. 2019/34/E/ST2/00369,
while K.{\.Z} acknowledges the ERC Advanced Grant {\sl Tatypic} number 101142236.

\bibliographystyle{unsrt}

\bibliography{bibliography}
\clearpage
\onecolumngrid
\appendix

\section{Distinguishing four-qudit AME states using invariants of local unitary (LU) equivalence
\label{app:LU_Invariants}
}
As all subsystems of AME states are maximally mixed it is difficult to distinguish one AME state or, equivalently its multi-unitary gate, from another based on known entanglement measures \cite{Rather_2022_Dual_Eq}. Fortunately for AME states composed of four qudits, Ref.~\cite{Rather_2023_AME_Eq} introduced invariants of LU equivalence that can distinguish AME states obtained using different constructions. Using these invariants of LU equivalence, we show that the AME states discussed in the main text obtained using the biunimodular construction are not LU equivalent to their graph state and stabilizer counterparts.

For a general bipartite unitary gate $U$ it can be shown that the spectrum of the following operator is invariant under LU transformation:
\begin{equation}
I(U)=(\mathcal{S}_{24} \otimes \mathbb{I})(U^{\dagger} \otimes U^{\dagger}) 
(\mathcal{S}_{24} \otimes \mathbb{I})(U \otimes U),
\end{equation}
where $\mathcal{S}_{24}$ is the SWAP gate between qudits 2 and 4; $\mathcal{S} \ket{i} \ket{j}=\ket{j}\ket{i}$ for all $i,j=0,1,\cdots,d-1$. It is easy to verify that the spectrum of $I(U)$ remains invariant under LU transformations on $U$; $(u_1 \otimes u_2)U (v_1\otimes v_2)$, where $u_i$ and $v_i$ are single-qudit gates. If $U$ and $U'$ are LU equivalent; $U'=(u_1 \otimes u_2)U(v_1\otimes v_2)$, then their corresponding four-qudit states are also LU equivalent; $\ket{U'}=(u_1\otimes u_2 \otimes v_1^T \otimes v_2^T) \ket{U}$.

It can be shown that the first moment of $I(U)$ is related to the operator entanglement of $U$ which is a well-known LU invariant \cite{Rather_2023_AME_Eq}. The first moment is maximized by all multi-unitary gates, therefore it cannot distinguish one multi-unitary gate from another. However, the second moment of $I(U)$; $\Tr[I(U)^2]$, turns out to be very useful for distinguishing multi-unitary gates belonging to different LU classes.

\begin{table}[]
    \centering
    \begin{tabular}{c||c||c}
 \hline
 Local dimension & AME state & LU invariant; $\Tr[I(\mathcal{U})^2]$ \\
 \hline
   & Graph state; Fig.~(\ref{fig:AME44}) & 256 \\
  4  & Minimal support state & 256 \\
   & Non-graph state; Eq.~(\ref{eq:U_AME(4,4)_F2}) & 64 \\
\hline
   & -- & -- \\
 6   & -- & -- \\
   & Non-graph state; Eq.~(\ref{eq:U_AME_4_6_F_23}) & 171 \\
\hline
   & Graph state & 4096 \\
 8   & Minimal support state & 1744 \\
   & Non-graph state: Eq.~(\ref{eq:U_AME_4_8_F_2}) & 314 \\
 \hline
 \hline
\end{tabular}
    \caption{The invariant of LU equivalence; $\Tr[I(\mathcal{U})^2]$, is evaluated for both graph and non-graph states. For a given local dimension $d$, different values of the invariant  illustrate that given AME states are not LU equivalent.}
    \label{Tab:AME_LU_Eq}
\end{table}

For multi-unitary gates $\mathcal{U}_1$ and $\mathcal{U}_2$ if $\Tr[I(\mathcal{U}_1)^2] \neq \Tr[I(\mathcal{U}_2)^2]$, then $\mathcal{U}_1$ and $\mathcal{U}_2$ are not LU equivalent. However, if $\Tr[I(\mathcal{U}_1)^2] = \Tr[I(\mathcal{U}_2)^2]$ then $\mathcal{U}_1$ and $\mathcal{U}_2$ may or may not be LU equivalent. For a given local dimension $d$, we compute $\Tr[I(\mathcal{U})^2]$ for multi-unitary gates corresponding to both graph and non-graph AME states as shown in Table \ref{Tab:AME_LU_Eq}. For local dimension $d=4$, the invariant takes different values for graph and non-graph states illustrating that these are not LU equivalent. However the invariant for graph and minimal support AME states is the same indicating that these may or may not be LU equivalent. Computing higher moments such as $\Tr[I(\mathcal{U})^4]$ surprisingly shows that these are in fact not LU equivalent. Interestingly, for local dimension $d=8$, the invariant takes different values in all three cases illustrating that these are not LU equivalent. For a given graph state in local dimension $d$, such as AME(4,4) shown in Fig.~(\ref{fig:AME44}), the corresponding multi-unitary gate is obtained by reshaping the given graph state (which is a $d^4$-dimensional unit vector) into a $d^2 \otimes d^2$ matrix and scaling it with a factor $d$ to preserve unitarity. For minimal support AME states the corresponding multi-unitary gates are simple permutation gates obtained from combinatorial designs such as orthogonal Latin squares \cite{Goyeneche2015}.

\section{Quantum circuits for non-stabilizer AME states of four ququarts; AME(4,4)
\label{app:AME_4_4_8_qubit_circ}
}
In this section, we provide further details about the AME states of four ququarts; AME(4,4), discussed in the main text along with some other new examples or equivalent AME states.
\subsection{Eight-qubit circuit involving 4-dimensional Fourier gate; $F_4$}
The construction of AME(4,4) states discussed in the main text uses unimodular vectors that are biunimodular with respect to $(F_2 \otimes F_2)^{\otimes 2}$. Here we illustrate that it is also possible to obtain AME(4,4) states using unimodular vectors that are biunimodular with respect to $F_4 \otimes F_4$. Let $\ket{\Lambda_4}$ be a biunimodular vector with respect to $(F_4 \otimes F_4)$ satisfying Eqs.~(\ref{eq:auto_corr}) and ~(\ref{eq:cross_auto_corr}), then the following unitary:
\begin{equation}
    \mathcal{U}[\Lambda_4]=CZ_4(F_4 \otimes F_4) D[\Lambda](F_4^{\dagger} \otimes F_4^{\dagger})CZ_4
    \label{eq:AME(4,4) v1},
\end{equation}
is a multi-unitary gate. An example of such a $\ket{\Lambda_4}$ that leads to a multi-unitary gate is given by

\begin{equation}
    \ket{\Lambda_4}=[ 1  , 1 , 1 ,-1  ,1 ,-1, -1, -1 , 1,  1 , 1, -1 ,-1  ,1,  1 , 1].
\end{equation}

\begin{figure*}
\begin{center}
\begin{quantikz}
\ket{0} &\gate{H}&\ctrl{4}&&&& \gategroup[4,steps=3,style={inner
sep=6pt, dashed,color=blue}]{$CZ_4$} & &
\ctrl{3}& \gate[2]{F_4^{\dagger}} & \ctrl{1} \gategroup[4,steps=6,style={inner
sep=6pt, dashed,color=red}]{$D[\Lambda_4]$} &&&&&& \gate[2]{F_4} & \gategroup[4,steps=3,style={inner
sep=6pt, dashed,color=blue}]{$CZ_4$} & & \ctrl{3} &\\[-0.5cm]
\ket{0} &\gate{H}&&\ctrl{4}&&&
\ctrl{2}&\ctrl{1}&& & \control{}&\gate{Z}&\ctrl{2}&\gate{X}&\ctrl{2}&\gate{X} & & \ctrl{2} & \ctrl{1} &&\\[-0.4cm]
\ket{0} &\gate{H}&&&\ctrl{4}&&&\control{}& & \gate[2]{F_4^{\dagger}} &  \gate{X}&  & \control{}  & \gate{X} & \control{} &  & \gate[2]{F_4} & & \control{} &&\\[-0.5cm]
\ket{0} &\gate{H}&&&&\ctrl{4}&\gate{S}&& \control{}&&   \gate{X}&  & \control{} & \gate{X} & \control{} &  & & \gate{S}&  & \control{}&\\
\ket{0} & & \targ{}  & &  & & & & & & & &&&&&&&&&\\[-0.4cm]
\ket{0} & &  & \targ{} & & &  & & & & & &&&&&&&&&\\[-0.4cm]
\ket{0} & & &  & \targ{} & & & & & & & &&&&&&&&&\\[-0.4cm]
\ket{0} & & &  &  & \targ{} & & & & & & &&&&&&&&&
\end{quantikz}    
\end{center}
\caption{Quantum circuit for AME state of four ququarts obtained from the multi-unitary gate given in Eq.~\eqref{eq:AME(4,4) v1}. Each ququart; $d=4$ is encoded into two qubits and the $CZ_4$ and $D[\Lambda_4]$ gates are decomposed in terms of single-, two-, and three-qubit gates.} 
\label{fig:AME_4_4_lam_4}
\end{figure*}
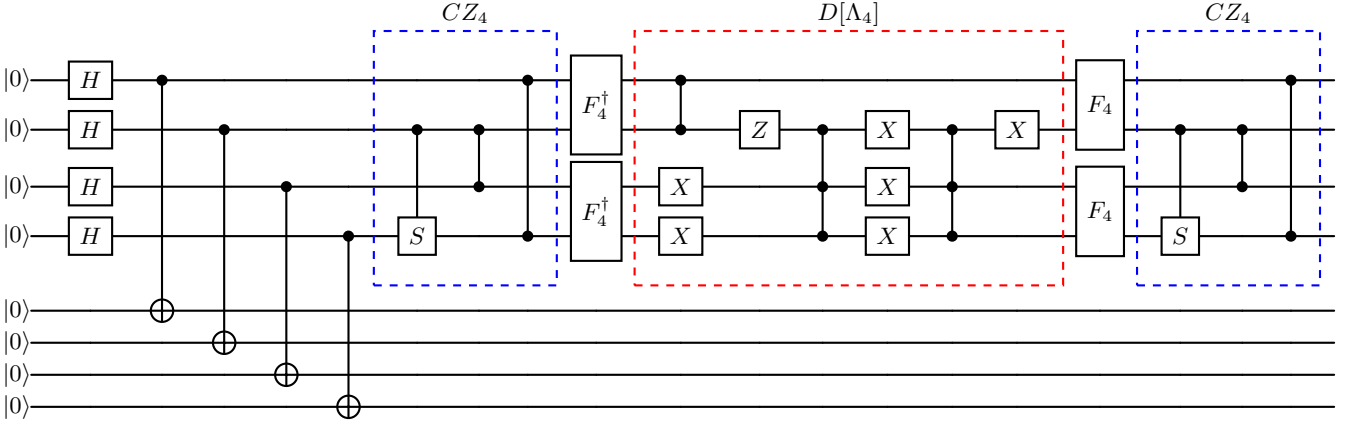

Both $CZ_4$ and $D[\Lambda_4]$ can be decomposed in terms of single- and two-qubit gates. A complete quantum circuit for obtaining AME(4,4) state using the multi-unitary gate $\mathcal{U}[\Lambda_4 ]$ is shown in Fig.~\ref{fig:AME_4_4_lam_4}. 

 \subsection{Another representation of the non-graph AME$(4,4)$ state}
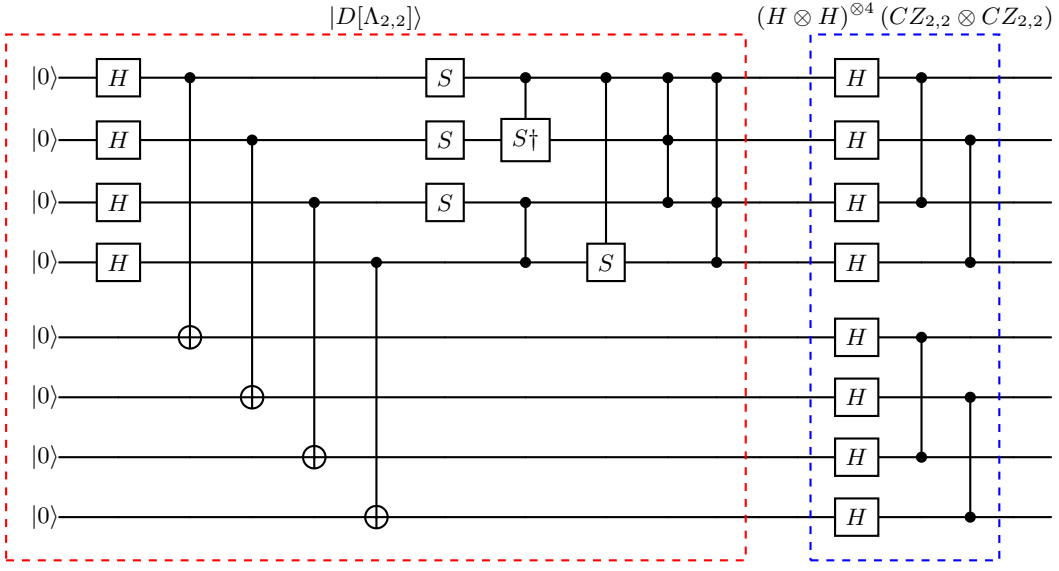
\begin{figure*}
\begin{center}
\begin{quantikz}
\gategroup[8,steps=11,style={inner
sep=6pt, dashed,color=red}]{$\ket{D[\Lambda_{2,2}]}$}  \ket{0} & \gate{H} & \ctrl{4} & & & 
& \gate{S} & \ctrl{1} & \ctrl{3}& \ctrl{2} & \ctrl{3} & & & \gate{H} \gategroup[8,steps=3,style={inner
sep=6pt, dashed,color=blue}] {$(H \otimes H)^{\otimes 4}\,(CZ_{2,2} \otimes CZ_{2,2})$} &  \ctrl{2} &  &   & \\[-0.2cm]

\ket{0}  & \gate{H} & & \ctrl{4} & & & 
   \gate{S} & \gate{S\dagger} & & \control{} & & & & \gate{H} &  & \ctrl{2}  & &  \\[-0.2cm]

\ket{0}  & \gate{H} & & & \ctrl{4} & &   \gate{S} & \ctrl{1} &  & \control{}  &  \control{} & & &  \gate{H} & \control{} & & & \\[-0.2cm]

\ket{0}  & \gate{H} & & & & \ctrl{4} &  &   \control{} &  \gate{S}  &  & \control{}  & & & \gate{H} &  & \control{} & &  \\

\ket{0}  & & \targ{}  & &  & & & & & & & & & \gate{H} & \ctrl{2} & & &\\[-0.2cm]

\ket{0}  & &  & \targ{} & & &  & & & & & & & \gate{H} & & \ctrl{2} & & \\[-0.2cm]

\ket{0}  & & &  & \targ{} & & & & & & & & & \gate{H} & \control{}
& & & \\[-0.2cm]

\ket{0}  & & &  &  & \targ{} & & & & & & & & \gate{H} & & \control{} & &
\end{quantikz}    
\end{center}
\caption{Eight-qubit quantum circuit for implementing the AME$(4,4)$ state given in Eq.~(\ref{eq:ket_AME_4_4_F2}).}
\label{fig:ket_AME_4_4_F2}
\end{figure*}

 Here we provide equivalent representations of the non-graph AME$(4,4)$ state discussed in the main text.
For operators $A_1,A_2,$ and $A_3$ on $\mathbb{C}^d$, the following vectorization identity holds;
\begin{equation}
  \ket{A_1A_2A_3}=(A_1\otimes A_3^T) \ket{A_2}. 
\end{equation}
Using the above vectorization identity in Eq.~(\ref{eq:U_AME(4,4)_F2}) we obtain the following equation:
\begin{align}
\label{eq:ket_AME_4_4_F2}
    \ket{\mathcal{U}[\Lambda_{2,2}]} & =\left(CZ_{2,2}(F_2\otimes F_2)^{\otimes 2} \otimes CZ_{2,2}(F_2\otimes F_2)^{\otimes 2}\right) \ket{D[\Lambda_{2,2}]}, \\ \nonumber 
    & =(CZ_{2,2} \otimes CZ_{2,2}) (F_2 \otimes F_2)^{\otimes 4}\ket{D[\Lambda_{2,2}]},
\end{align}

where $F_2 \equiv H$ is the single-qubit Hadamard gate and $\ket{D[\Lambda_{2,2}]}$ is the vectorization of $D[\Lambda_{2,2}]$. The $CZ_{2,2}$ gate can be easily decomposed in terms of the two-qubit $CZ$ gate [Cf. Eq.~(\ref{eq:CZ_2_2_decom})]. Using the decompositions of $CZ_{2,2}$ and $D[\Lambda_{2,2}]$ gates [Cf. Eq.~(\ref{eq:Dlam_2_2_decom})], a quantum circuit to implement Eq.~(\ref{eq:ket_AME_4_4_F2}) using qubits is shown in Fig.~(\ref{fig:ket_AME_4_4_F2}).

\section{Decomposition of the diagonal gate appearing in the an AME state of four quhexes; AME(4,6)} \label{app:qubit_circuit}

In this section, we provide a decomposition of the diagonal gate appearing in the construction of the AME$(4,6)$ state discussed in Section~\ref{sec:AME46_qubits}. For clarity, we restate the diagonal gate:
\begin{equation}
D[\Lambda_{2,3}]=\text{Diag}[ 1, \overline{\omega_3}, \overline{\omega_3}, \overline{\omega_3}, \overline{\omega_3}, 1, 1, 1, \omega_3, 1, \omega_3, 1, 1, \omega_3, 1, \omega_3
                , 1, 1, 1, \omega_3, 1, \overline{\omega_3}, \omega_3, \overline{\omega_3}, \overline{\omega_3}, \omega_3, \omega_3, \overline{\omega_3}, 1, 1, \omega_3, \omega_3, \overline{\omega_3}, 1, 1, \overline{\omega_3}],
\label{eq:diagonal_gate_qubits_appendix}
\end{equation}
where $\omega_3 = \exp(2\pi i/3)$ and $\overline{\omega_3}$ is its complex conjugate. To implement this gate using qubits, we first need to map the qudits levels to qubits. For this, we use a standard qudit-to-qubit mapping: 
\begin{equation}
    \ket{0}\rightarrow\ket{000}, \quad \ket{1}\rightarrow\ket{001}, \quad 
    \ket{2}\rightarrow\ket{010}, \quad 
    \ket{3}\rightarrow\ket{011}, \quad 
    \ket{4}\rightarrow\ket{100}, \quad 
    \ket{5}\rightarrow\ket{101}. \quad 
\end{equation}
The states $\ket{110}$ and $\ket{111}$ are not included in this mapping. Therefore, the unitary transformation must act trivially on these states. 

\begin{figure}
    \centering
    \includegraphics[width= 1\linewidth]{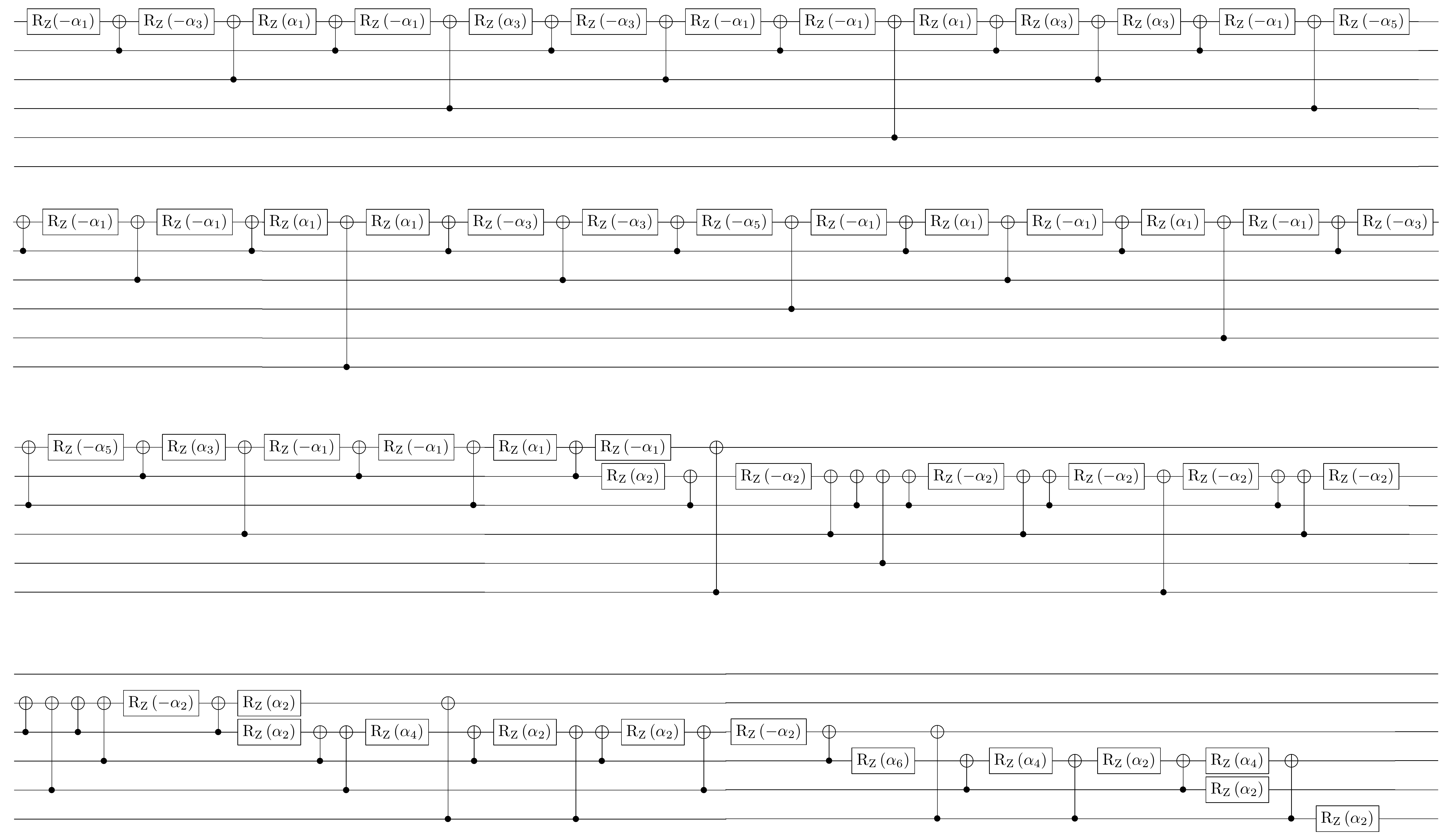}
    \caption{Diagonal gate decomposition for an AME state of four quhexes: AME(4,6). Qubit quantum circuit for the diagonal gate in Eq.~\eqref{eq:diagonal_gate_qubits}. Each wire corresponds to a qubit, and the angles in the $Z-$rotations are given by $\alpha_j := j\frac{\pi}{24}$, for $j\in \{1,2,3,4,6\}$. For constructing the qubit circuit, we mapped each qudit to three qubits. }
    \label{fig:diagonal_gate_qubits}
\end{figure}

The resulting circuit according to this mapping is given in Fig.~\ref{fig:diagonal_gate_qubits}. It can be prepared using only controlled-not gates and rotations along the $Z$-axis. Despite its large depth, the circuit has a certain structure. For example, all angle rotations are multiples of $\frac{\pi}{24}$. We leave it as future work to explore whether the depth of this circuit can be reduced with improved qudit-to-qubit mapping strategies or with other unitary compilation procedures.

\section{Decomposition of the diagonal gate for AME state of four quocts; AME(4,8) \label{app:Diag_2_2_2_qubit_decom}
}

\begin{figure}
    \centering
    \includegraphics[width=1\linewidth]{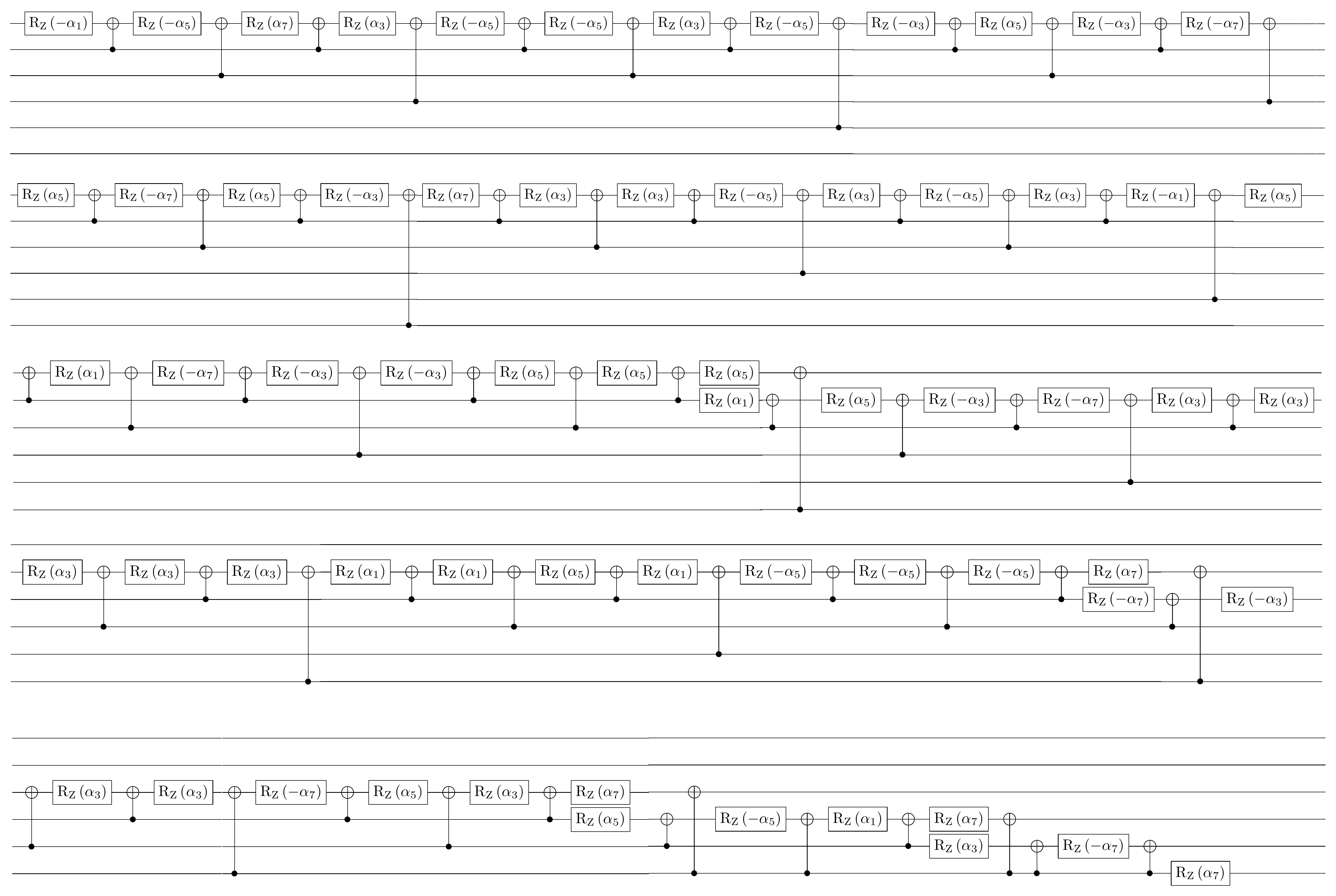}
    \caption{Diagonal gate decomposition for an AME state of four quocts: AME(4,8). Qubit quantum circuit for the diagonal gate given in Eq.~\eqref{eq:D_lam_2_2_2_app}. Each wire corresponds to a qubit, and the angles in the $Z-$rotations are given by $\alpha_j := j\frac{\pi}{32}$, for $j\in \{1,3,5,7\}$. For constructing the qubit circuit, we mapped each 8-dimensional qudit to three qubits. The circuit should be read from from left to right and top to bottom.}
    \label{fig:diagonal_circuit_222}
\end{figure}

Here we provide the circuit decomposition for the diagonal gate required in the construction of the AME$(4,8)$ state, discussed in Sec.~\ref{sec:ame(4,8)}. The diagonal gate $D[{\Lambda_{2,2,2}}]$ appearing in the constructions of AME(4,8) state; see Fig.~(\ref{fig:AME_4_8_circ}), is given by 
\begin{align}
 \label{eq:D_lam_2_2_2_app}
D[{\Lambda_{2,2,2}}]= \text{Diag}[& +,  +,  i,  +, -i,  -,   i,  +,  -i,  -,  -, -, i, -i,  +,   -, 
                             i,  i, -i, -i,  +, -i,  -i,  i,  -i,  +,  -, i, -,  +,  i,   -, \\ \nonumber
                           & -,  +, -,   i,  -,  +,   i, -i,   +,  +, -i, -, -,  +,  i,  -i, 
                             -,  -, -,   -,  -,  +,  -i,  +,   -,  -,  -, -,  i, i, -i,  -],
\end{align}
where $\mp$'s denote $\mp 1$'s. Encoding each quoct; $d=8$, into three qubits the circuit decomposition of the above diagonal gate using single- and two-qubit CNOT gates is shown in Fig.~\ref{fig:diagonal_circuit_222}. Again, we see that the angles in the $Z-$rotations are all multiples of a fraction, in this case, $\frac{\pi}{32}$.

\section{Certification of genuine multipartite entanglement}\label{app:entanglement}

In this section, based on Theorem~\ref{th:genuine_entanglement} from Ref.~\cite{malik2016multi}, we provide an estimation for the minimum fidelity for an AME$(4,d)$ state to be entangled in a $d^4$-dimensional space, namely $\mathbb{C}^d\otimes\mathbb{C}^d\otimes\mathbb{C}^d\otimes\mathbb{C}^d$. This implies that the experimental state satisfying this condition cannot be decomposed into entangled states of smaller dimensions. In other words, we want to derive the minimum fidelity that we can accept to ensure that our state lies outside the convex hull of states that can be decomposed into other states where at least one of the particles has local dimension is $d-1$. Let us denote such a set by $(d,d,d,d-1)$.

The ideal AME$(4,d)$ state is a pure state $\ket{\psi}$; however, due to noise, when implemented in a quantum device, the resulting state becomes a mixed one, which we denote by $\rhoexp$. We can obtain the fidelity with the ideal state as $\Fexp := \Tr(\rhoexp \ket{\psi}\bra{\psi})$. The maximum fidelity that a $(d, d, d, d-1)$-state can achieve with respect of the AME$(4,d)$ is given by 
\begin{equation}
   \Fmax:= \max_{\sigma\in (d,d,d,d-1)} \Tr (\sigma  \ket{\psi}\bra{\psi}).
   \label{eq:F_max_app}
\end{equation}
If $\Fexp> \Fmax$ then the experimental state is certified to be genuinely entangled in $(\mathbb{C}^d)^{\otimes 4}$. The set of $(d,d,d, d-1)$ states is convex, which means that the maximum in Eq.~\eqref{eq:F_max_app} can always be achieved by a pure state:
\begin{equation}
\begin{split}
    \Fmax = \max_{\ket{\phi}\in (d,d,d,d-1)} |\braket{\psi|\phi}|^2 = \max \Big\{ &\max_{\ket{\phi}\in (d,d,d,d-1)} |\braket{\psi|\phi}|^2 , \max_{\ket{\phi}\in (d,d,d-1,d)} |\braket{\psi|\phi}|^2 , \\&\max_{\ket{\phi}\in (d,d-1,d,d)} |\braket{\psi|\phi}|^2 , \max_{\ket{\phi}\in (d-1,d,d,d)} |\braket{\psi|\phi}|^2 \Big\}.
    \label{eq:F_max}
\end{split}
\end{equation}
For each of the four terms in the above expression, we are going to use the bound on a vector of fixed rank $(x,y,z,t)$
\begin{equation}
\begin{split}
    \max_{\ket{\phi}\in (x,y,z,t)} |\braket{\psi|\phi}|^2 \leq \min \Big\{  &\max_{{\rm r}(\Tr_{234}\ket{\phi}\bra{\phi}) = x} |\braket{\psi|\phi}|^2 , \max_{{\rm r}(\Tr_{134}\ket{\phi}\bra{\phi}) = y} |\braket{\psi|\phi}|^2, \\ &\max_{{\rm r}
    (\Tr_{124}\ket{\phi}\bra{\phi}) = z} |\braket{\psi|\phi}|^2, \max_{{\rm r}(\Tr_{123}\ket{\phi}\bra{\phi}) = t} |\braket{\psi|\phi}|^2 \Big\}.
    \label{eq:bound_fixed_rank}
\end{split}
\end{equation}
If a state admits Schmidt decomposition across a bipartition $A|\bar A$, given by $\ket\psi = \sum_{i=0}^{r-1}\lambda_i \ket{v_A^i}\otimes \ket{v^i_{\bar A}}$ then, assuming ordered Schmidt coefficients, we can compute the maximal possible overlap with a state of bounded rank for each term in~\eqref{eq:bound_fixed_rank} as
\begin{equation}
    \max_{rank(\Tr_{\bar A}\ket{\phi}\bra{\phi}) = x} |\braket{\psi|\phi}|^2 = \sum_{i =0}^{x-1} \lambda_{i}^2.
    \label{eq:Schmidt_coeff_theorem}
\end{equation}
Hence, we need to find the Schmidt coefficients for all bipartitions of the AME state. By definition of absolutely maximally entangled states, the Schmidt coefficients across bipartitions $1|123$, $2|134$, $3|124$, and $4|123$ are all the same, and they are given by the collection of $d$ elements $\{\frac{1} {\sqrt{d}},\frac{1} {\sqrt{d}}, ..., \frac{1} {\sqrt{d}} \}$. By introducing the bound from Eq.~\eqref{eq:bound_fixed_rank} into Eq.~\eqref{eq:F_max} and using the result in Eq.~\eqref{eq:Schmidt_coeff_theorem}, we eventually arrive at the value of the maximum fidelity. This value imposes a lower bound in the experimental fidelity required to achieve genuine multipartite entanglement in a space of dimension $d^4$, namely
\begin{equation}
    \Fmax = \frac {d-1}{d} \implies \Fexp\geq \frac{d}{d-1}.
    \label{eq:max_fidelity_AME44}
\end{equation}

\end{document}